\documentclass{article}
\usepackage[utf8]{inputenc}
\usepackage{geometry}
\usepackage{auxlib}

\usepackage{authblk}

\def\srma/{SRMA}
\def\srcp/{SRCP}
\def\scma/{SCMA}
\usepackage{biblatex}
\title{Min-max Submodular Ranking for Multiple Agents\thanks{All authors (ordered alphabetically) have equal contributions and are corresponding authors.}}
\date{}

\author[1]{Qingyun Chen}
\author[1]{Sungjin Im}
\author[2]{Benjamin Moseley}
\author[3,4\thanks{This work was done when the author was a student at Zhejiang University.}]{Chenyang Xu}
\author[5\thanks{This work was done when the author visited Carnegie Mellon University.} ]{Ruilong Zhang}

\affil[1]{\footnotesize Electrical Engineering and Computer Science, University of California at Merced}
\affil[2]{\footnotesize Tepper School of Business, Carnegie Mellon University}
\affil[3]{\footnotesize Software Engineering Institute, East China Normal University}
\affil[4]{\footnotesize College of Computer Science, Zhejiang University}
\affil[5]{\footnotesize Department of Computer Science, City University of Hong Kong}
\affil[ ]{\texttt{qchen41@ucmerced.edu, sim3@ucmerced.edu, moseleyb@andrew.cmu.edu, xcy1995@zju.edu.cn, ruilzhang4-c@my.cityu.edu.hk}}

\begin{document}

% \keywords{}

\maketitle

\begin{abstract}
In the submodular ranking (SR) problem, the input consists of a set of submodular functions defined on a ground set of elements. The goal is to order elements for all the functions to have value above a certain threshold as soon on average as possible, assuming we choose one element per time. 
The problem is flexible enough to capture various applications in machine learning, including decision trees. 

This paper considers the min-max version of SR where multiple instances share the ground set. With the view of each instance being associated with an agent, the min-max problem is to order the common elements to minimize the maximum objective of all agents---thus, finding a fair solution for all agents. We give approximation algorithms for this problem and demonstrate their effectiveness in the application of finding a decision tree for multiple agents. 
\end{abstract}

\clearpage

\section{Introduction}
    \label{sec:intro}

% \qingyun{To do note}
% \sungjin{To do note}
% \ben{To do note}
% \chenyang{To do note}
% \ruilong{To do note}

The  submodular ranking (SR) problem was proposed by \cite{DBLP:conf/soda/AzarG11}.
The problem includes a ground set of elements $[n]$, and a collection of $m$ monotone submodular\footnote{A function $f: 2^{[n]} \rightarrow \R$ is submodular if for all $A, B \subseteq [n]$ we have $f(A) + f(B) \geq f(A \cap B) + f(A \cup B)$. The function is monotone if $f(A) \leq f(B)$ for all $A \subseteq B$.} set functions $f_1,\ldots,f_m$ defined over $[n]$, i.e., $f_j:2^{[n]}\to \R_{\geq 0}$ for all $j\in[m]$.
Each function $f_j$ is additionally associated with a positive weight $w_j\in \R$.
Given a permutation $\pi$ of the ground set of elements, the cover time of a function $f_j$ is defined to be the minimal number of elements in the prefix of $\pi$ which forms a set such that the corresponding function value is larger than a unit threshold value\footnote{If the threshold is not unit-valued, one can easily obtain an equivalent instance by normalizing the corresponding function.}.
The goal is to find a permutation of the elements such that the weighted average cover time is minimized.

The SR problem has a natural interpretation. Suppose there are $m$ types of clients who are characterized by their utility function $\{f_j\}_{j \in [m]}$, and we have to satisfy a client sampled from a known probability distribution $\{w_j\}_{j \in [m]}$. Without knowing her type, we need to sequentially present items to satisfy her, corresponding to her utility $f_j$ having a value above a threshold. Thus, the goal becomes finding the best ordering to minimize the average number of items shown until satisfying the randomly chosen client. 

In the \emph{min-max} submodular ranking for multiple agents, we are in the scenario where there are $k$ SR instances sharing the common ground set of elements, and the goal is to find a common ordering that minimizes the maximum objective of all instances. Using the above analogy, suppose there are $k$ different groups of clients. We have $k$ clients to satisfy, one sample from each group. Then, we want to satisfy \emph{all} the $k$ clients as early as possible. In other words, we want to commit to an ordering that satisfies all groups fairly---formalized in a min-max objective. 

\subsection{Problem Definition}
    \label{sec:prelim}

The problem of min-max submodular ranking for multiple agents (\srma/) is formally defined as follows. 
An instance of this problem consists of a ground set $U:=[n]$, and a set of $k$ agents $A:=\set{a_1,\ldots,a_k}$.
Every agent $a_i$, where $i\in[k]$, is associated with a collection of $m$ monotone submodular set functions $f_1^{(i)},\ldots,f_m^{(i)}$. It is the case that $f^{(i)}_j:2^{[n]}\to[0,1]$ with $\fij(U)=1$ for all $i\in[k],j\in[m]$.
In addition, every function $\fij$ is associated with a weight $w^{(i)}_j>0$ for all $i\in[k],j\in[m]$.

Given a permutation $\pi$ of the ground elements, the cover time $\cov(\fij,\pi)$ of $\fij$ in $\pi$ is defined as the smallest index $t$ such that the function $\fij$ has value $1$ for the first $t$ elements in the given ordering. 
The goal is to find a permutation of the ground elements that minimizes the maximum total weighted cover time among all agents, i.e., finding a permutation $\pi$ such that $\max_{i\in[k]}\left\{\sum_{j=1}^{m}\wij\cdot\cov(\fij,\pi)\right\}$ is minimized.
We assume that the minimum non-zero marginal value of all $m\cdot k$ functions is $\epsilon>0$, i.e., for any $S\subseteq S'$, $\fij(S')>\fij(S)$ implies that $\fij(S')-\fij(S)\geq\epsilon>0$ for all $i\in[k],j\in[m]$.
Without loss of generality, we assume that any $w^{(i)}_j\geq 1$ and let $W:=\max_{i\in[k]}\sum_{j=1}^{m}\wij$ be the maximum total weight among all agents.

\subsection{Applications}

In addition to the aforementioned natural interpretation, we discuss two concrete applications below in detail.

\paragraph{Optimal Decision Tree (ODT) with Multiple Probability Distributions.}
In (the non-adaptive) optimal decision tree problem with multiple probability distributions, we are given $k$ probability distributions $\{p^{(i)}_j\}_{j \in [m]}$, $i \in [k]$ 
over $m$ hypotheses  and a set $U := [n]$ of binary tests. 
There is exactly one unknown hypothesis $\widetilde{j}^{(i)}$ drawn from $\{p^{(i)}_j\}_{j \in [m]}$ for each $i \in [k]$. The outcome of each test $e \in U$ is a partition of $m$ hypotheses. Our goal is to find a permutation of $U$ that minimizes the expected number of tests to identify \emph{all} the $k$ sampled hypotheses $\widetilde{j}^{(i)}, i \in [k]$. 

This problem generalizes the traditional optimal decision tree problem which assumes $k = 1$. The problem has the following motivation. Suppose there are $m$ possible diseases and their occurrence rate varies depending on the demographics. A diagnosis process should typically be unified and the process should be fair to all demographic groups. If we adopt the min-max fairness notion, the goal then becomes to find a common diagnostic protocol that successfully diagnoses the disease to minimize the maximum diagnosis time for any of the $k$ groups.  The groups are referred to as agents in our problem. 

%the maximum expected number of tests over all agents to rule out the unknown hypothesis $j^*$. The goal is to find a permutation of $U$ that minimizes the maximum expected number of tests over all agents to rule out the unknown hypothesis

%In Optimal Decision Tree problem for Multiple Agents, for each agent $i$, we are given a set of $m$ hypotheses with a probability distribution $\{p^{(i)}_j\}_{j \in [m]}$ with a weight $w^{(i)}_j$ and a set $U := [n]$ of binary tests. There is only one unknown hypothesis $j^*$ drawn from $\{p^{(i)}_j\}_{j \in [m]}$. The outcome of each test $e \in U$ is a partition of $m$ hypotheses. Our goal is to find a permutation of $U$ that minimizes the maximum expected number of tests over all agents to rule out the unknown hypothesis $j^*$.

As observed in \cite{jia2019optimal}, the optimal decision tree is a special case of submodular ranking. To see this, fix agent $a_i$. For notational simplicity we drop $a_i$. %Let $T(e)$ be the set of hypotheses whose outcome is negative by test $e$.
Let $T_j(e)$ be the set of hypotheses that have an outcome different from $j$ for test $e$. 
For each hypothesis $j$, we define a monotone submodular function $f_j: 2^U \to [0,1]$ with a weight $p_j$ as follows:
$$ f_j(S) = \bigg|\bigcup_{e \in S} T_j(e)\bigg| \cdot \frac{1}{m - 1} $$
which is the fraction of hypotheses other than $j$ that have been ruled out by a set of tests $S$. Then, hypothesis $\widetilde{j}$ can be identified if and only if $f_{\widetilde{j}}(S) = 1$.

\paragraph{Web Search Ranking with Multiple User Groups.}
Besides fair decision tree, our problem also captures several practical applications, as discussed in \cite{DBLP:conf/soda/AzarG11}.
Here, we describe the application of web search ranking with multiple user groups.
A user group is drawn from a given distribution defined over $k$ user groups.
We would like to display the search results sequentially from top to bottom.
We assume that each user browses search results from top to bottom.
Each user is satisfied when she has found sufficient web pages relevant to the search, and the satisfaction time corresponds to the number of search results she checked.
The satisfaction time of a user group is defined as the total satisfied time of all users in this group.
The min-max objective ensures fairness among different groups of users.

\subsection{Our Contributions}
\label{sec:contri}

To our knowledge, this is the first work that studies submodular ranking, or its related problems, such as optimal decision trees, in the presence of multiple agents (or groups).
Finding an order to satisfy all agents equally is critical to be fair to them. This is because the optimum ordering can be highly different for each agent.

% Finding an ordering to satisfy all agents equally is critical to be fair to them. 
% This is because the optimum ordering can be highly different for each group.
We first consider a natural adaptation of the normalized greedy algorithm \cite{DBLP:conf/soda/AzarG11}, which is the best algorithm for submodular ranking.
We show that the adaptation is $O(k \ln( 1/ \epsilon))$-approximation and has an $\Omega(\sqrt{k})$ lower bound on the approximation ratio.

%We show that it does not have a good approximation ratio.
%Namely, we show that the approximation ratio is $O(k \ln( 1/ \epsilon))$-approximation and $\Omega(\sqrt{k})$ (See \cref{sec:ng} for details). 
%Intuitively, an algorithm needs to balance across the groups to ensure fairness and greedy fails to do this well. 

% We first consider  a natural adaptation of the normalized greedy algorithm \cite{DBLP:conf/soda/AzarG11} for submodular ranking and  show that it is an $O(k \ln( 1/ \epsilon))$-approximation. But, for the min-max optimization, an algorithm needs to balance across the agents to ensure fairness.

%The $\Omega(\sqrt{k})$ lower bound suggests a critical need to balance the progress of agents.
% The $\Omega(\sqrt{k})$ lower bound suggests that the previous normalization techniques 
To get rid of the polynomial dependence on $k$, we then develop a new algorithm, which we term \emph{balanced adaptive greedy}. 
Our new algorithm overcomes the polynomial dependence on $k$ by carefully balancing agents in each phase.
To illustrate the idea,  for simplicity, assume all weights associated with the functions are 1 in this paragraph. We iteratively set checkpoints: in the $p$-th iteration, we ensure that we satisfy all except about $m / 2^p$ functions for each agent. By balancing the progress among different agents, we obtain a $O( \ln(1 / \epsilon)\log(\min\{n,m\}) \log k )$ approximation
(\cref{thm:main}), where $n$ is the number of elements to be ordered.
When $\min\{n,m\} = o(2^k)$, the balanced adaptive greedy algorithm has a better approximation ratio than the natural adaptation of the normalized greedy algorithm.
For the case that $\min\{n,m\} = \Omega(2^k)$, we can simply run both of the two algorithms and pick the better solution which yields an algorithm that is always better than the natural normalized greedy algorithm.
%We hope our idea of balancing agents can be used in other min-max multiagent settings.
%Note that one can easily get a better ratio $\min\{\gamma_1,\gamma_2\}$ by combining the natural normalized greedy algorithm and balanced adaptive greedy.

We complement our result by showing that it is NP-hard to approximate the problem within a factor of $O(\ln ( 1/ \epsilon) + \log k)$ (\cref{cor:hardness}). 
While we show a tight $\Theta(\log k)$-approximation for generalized min-sum set cover over multiple agents which is a special case of our problem (see \cref{thm:mssc}), reducing the gap in the general case is left as an open problem.
% we do not know tight approximation for the general case. Reducing the gap is left as an open problem.

We demonstrate that our algorithm outperforms other baseline algorithms for real-world data sets. This shows that the theory is predictive of practice. The experiments are performed for optimal decision trees, which is perhaps the most important application of submodular ranking.

\subsection{Related Works}

% In this work, we extend the problem of Submodular Ranking (SR) to more general case. 

\paragraph{Submodular Optimization.}
Submodularity commonly arises in various practical scenarios. It is best characterized by diminishing marginal gain, and it is a common phenomenon observed in all disciplines. Particularly in machine learning, submodularity is useful as a considerable number of problems can be cast into submodular optimizations, and (continuous extensions of) submodular functions can be used as regularization functions \cite{bach2013learning}. Due to the extensive literature on submodularity, we only discuss the most relevant work here. Given a monotone submodular function $f: 2^U \rightarrow \Z_{\geq 0}$, choosing a $S \subseteq U$ of minimum cardinality s.t. $f(S) \geq T$ for some target $T \geq 0$ admits a $O(\log T)$-approximation \cite{williamson2011design}, which can be achieved by iteratively choosing an element that increase the function value the most, i.e., an element with the maximum marginal gain. Or, if $f$ has a range $[0, 1]$,  $T = 1$ and the non-zero marginal increase is at least $\epsilon >0$, we can obtain an $O(\ln (1 / \epsilon))$-approximation.

\paragraph{Submodular Ranking.}
The submodular ranking problem was introduced by \cite{DBLP:conf/soda/AzarG11}.
In \cite{DBLP:conf/soda/AzarG11}, they gave an elegant greedy algorithm that achieves $O(\ln\frac{1}{\epsilon})$-approximation and provided an asymptotically matching lower bound.
The key idea was to renormalize the submodular functions over the course of the algorithm. That is, if $S$ is the elements chosen so far, we choose an element $e$ maximizing $\sum_{f\in \cF} w_f \cdot \frac{f(S \cup \{e\})-f(S)}{1 - f(S)}$, where $\cF$ is the set of uncovered functions. Thus, if a function $f_j$ is nearly covered, then the algorithm considers $f_j$  equally by renormalizing it by the residual to full coverage, i.e., $1 - f_j(S)$. Later, \cite{DBLP:journals/talg/ImNZ16} gave a simpler analysis of the same greedy algorithm and extended it to other settings involving metrics. Special cases of submodular ranking include min-sum set cover \cite{feige2004approximating} and generalized min-sum set cover \cite{azar2009multiple,bansal2021improved}. For stochastic extension of the problem, see \cite{DBLP:journals/talg/ImNZ16,agarwal2019stochastic,jia2019optimal}.%hellerstein2021tight}.

\paragraph{Optimal Decision Tree.} 
The optimal decision tree problem (with one distribution) has been extensively studied. The best known approximation ratio for the problem is $O( \log m)$ \cite{gupta2017approximation} where $m$ is the number of hypotheses and it is asymptotically optimal unless P $=$ NP \cite{chakaravarthy2007decision}. As discussed above, it is shown in \cite{jia2019optimal} how the optimal decision tree problem is captured by the submodular ranking. For applications of optimal decision trees, see~\cite{dasgupta2004analysis}. 
Submodular ranking only captures non-adaptive optimal decision trees. For adaptive and noisy decision  trees, see \cite{golovin2010near,jia2019optimal}.

\paragraph{Fair Algorithms.} 
Fairness is an increasingly important criterion in various machine learning applications \cite{barocas2017fairness,chouldechova2020snapshot,mehrabi2021survey,DBLP:conf/nips/HalabiMNTT20,DBLP:conf/icml/AbernethyAKM0Z22}, yet certain fairness conditions cannot be satisfied simultaneously \cite{kleinberg2016inherent,corbett2017algorithmic}.
In this paper, we take the min-max fairness notion, which is simple yet widely accepted. For min-max, or max-min fairness, see \cite{radunovic2007unified}.
\section{Warm-up Algorithm: A Natural Adaptation of Normalized Greedy}
\label{sec:ng}

The normalized greedy algorithm~\cite{DBLP:conf/soda/AzarG11} obtains the best possible approximation ratio of $O(\ln \frac{1}{\epsilon})$ on the traditional (single-agent) submodular ranking problem, where $\epsilon$ is the minimum non-zero marginal value of functions. The algorithm picks the elements sequentially and the final element sequence gives a permutation of all elements. Each time, the algorithm chooses an element $e$ maximizing $\sum_{f\in \cF} w_f \cdot \frac{f(S \cup \{e\})-f(S)}{1 - f(S)}$, where $S$ is the elements chosen so far and $\cF$ is the set of uncovered functions.

For the multi-agent setting, there exists a natural adaptation of this algorithm: simply view all $m\cdot k$ functions as a large single-agent submodular ranking instance and run normalized greedy. For simplicity, refer to this adaption as algorithm NG.
The following shows that this algorithm has to lose a polynomial factor of $k$ on the approximation ratio.

% The shortfall of the algorithm is that it misjudges elements with low marginal contribution as less important, and thus, schedules them late. As one may expect, this turns to be crucial when many functions depend on a single element which has a low marginal contribution.

\begin{theorem}\label{thm:NG_Lower_bound}
The approximation ratio of algorithm NG is $\Omega(\sqrt{k})$, even when the sum $\sum_{j} w_j^{(i)}$ of function weights for each agent $i$ is the same.
\end{theorem}

\begin{proof}
Consider the following \srma/ instance. There are $k$ agents and each agent has at most $\sqrt{k}$ functions; $\sqrt k$ is assumed to be an integer. We first describe a weighted set cover instance and explain how the instance is mapped to an instance for our problem. We are given a ground set of $k + \sqrt{k}$ elements $E = \set{e_1, e_2, \ldots, e_{k+\sqrt{k}}}$ and a collection of singleton sets corresponding to the elements. 
Each agent $i \in [k - 1]$ has two sets, $\set{e_i}$ with weight $1+\delta$ and $\set{e_k}$ with weight $\sqrt{k} - 1-\delta$, where $\delta>0$ is a tiny value used for tie-breaking. The last agent $k$ is special and she has $\sqrt k$ sets, $\set{e_{k+1}}, \ldots, \set{e_{k+\sqrt{k}}}$, each with weight $1$.  Note that every agent has exactly the same total weight~$\sqrt{k}$. 
 
Each set is ``covered" when we choose the unique element in the singleton set. This results in an instance for our problem: for each set $\{e\}$ with weight $w$, we create a distinct 0-1 valued function $f$ with an equal weight $w$ such that $f(S) = 1$ if and only if $e \in S$. It is obvious to see that all created functions are submodular and have function values in $\{0, 1\}$. It is worth noting that $e_k$ is special and all functions of the largest weight $(\sqrt{k}-1)$ get covered simultaneously when $e_k$ is selected. 
 
%Because all functions originate from singleton sets, it is easy to see that algorithm NG behaves exactly the same as the greedy. Thus, in the following we will just refer to NANG---but we are showing a lower bound for both algorithms. 

Algorithm NG selects $e_k$ in the first step.
After that, the contribution of each element in $E\setminus\set{e_k}$ in each following step is the same, which is $\sqrt k - 1$. Thus, the permutation $\pi$ returned by algorithm NG is $(e_k,e_1,\ldots,e_{k-1},e_{k+1},\ldots,e_{k+\sqrt{k}})$. 
The objective for this ordering is at least $\Omega(k^{1.5})$ since until we choose $e_{k + \sqrt k / 2}$, the last agent $k$ has at least $\sqrt k / 2$ functions unsatisfied. 
%whose objective value is $\frac{\sqrt{k}}{2}(2k+\sqrt{k}+1)$.
However, another permutation $\pi'=(e_k,e_{k+1},\ldots,e_{k+\sqrt{k}},e_1,\ldots,e_{k-1})$ obtains an objective value of $O(k)$. This implies that the approximation ratio of the algorithm is $\Theta(\sqrt{k})$ and completes the proof.
\end{proof}

%Given an arbitrary instance $I$ of \srma/, we construct an instance $I_S$ of the Submodular Ranking problem.
%There is only one agent $a$ in $I_S$ and this agent controls $mk$ monotone submodular functions. 
%Now, we run the greedy algorithm proposed in \cite{DBLP:conf/soda/AzarG11} on instance $I_S$.

One can easily show that the approximation ratio of algorithm NG is $O(k\ln(\frac{1}{\epsilon}))$ by observing that the total cover time among all agents is at most $k$ times the maximum cover time.
The details can be found in the proof of the following theorem.

\begin{theorem}
Algorithm NG achieves $(4k\ln(\frac{1}{\epsilon})+8k)$-approximation.
\label{thm:ratio:NG}
\end{theorem}

\begin{proof}
%[Proof of \cref{thm:ratio:NG}]
%We show that the solution $\pi$ returned by the greedy algorithm is the desired solution.
%Use $\pi$ to denote the permutation returned by algorithm NG.
Consider any element permutation $\sigma$.
Let $F(I,\sigma)$ and $F(I_{\cup},\sigma)$ be the objective value of instance $I$ and $I_{\cup}$ when the permutation $\sigma$ is applied, where $I$ is an instance of \srma/ and $I_{\cup}$ is the stacked single-agent submodular ranking instance of $I$.
In other words, $F(I_{\cup},\sigma)=\sum_{i=1}^{k}\sum_{j=1}^{m}\wij\cdot\cov(\fij,\sigma)$ and $F(I,\sigma)=\max_{i\in[k]}\{\sum_{j=1}^{m}\wij\cdot\cov(\fij,\sigma)\}$.
Clearly, for any permutation $\sigma$, we have $F(I,\sigma)\leq F(I_{\cup},\sigma)$. %(this inequality holds for any permutation).

Use $\pi$ to denote the permutation returned by algorithm NG.
Let $\pi'$ and $\pi''$ be the optimal solution to $I$ and $I_{\cup}$, respectively.
Since $\pi'$ is a feasible solution to $I_{\cup}$, we have $F(I_{\cup},\pi'')\leq F(I_{\cup},\pi')$.
Thus, 
\begin{align}
\begin{split}
F(I_{\cup},\pi'')&\leq F(I_{\cup},\pi') \\
&=\sum_{i=1}^{k}\sum_{j=1}^{m}\wij\cdot\cov(\fij,\pi') \\
&\leq k\cdot\left( \max_{i\in[k]}\left\{\sum_{j=1}^{m}\wij\cdot\cov(\fij,\pi')\right\} \right) \\
&=k\cdot F(I,\pi') 
\label{equ:alg-trivial:opt}
\end{split}
\end{align}
By \cite{DBLP:conf/soda/AzarG11}, we know that $\pi$ is a $(4\ln(\frac{1}{\epsilon})+8)$-approximation solution to instance $I_{\cup}$, i.e., $F(I_{\cup},\pi)\leq (8+4\ln(\frac{1}{\epsilon}))\cdot F(I_{\cup},\pi'')$.
Since $F(I,\pi)\leq F(I_{\cup},\pi)$, we have $F(I,\pi)\leq (4k\ln(\frac{1}{\epsilon})+8k)\cdot F(I,\pi')$ by \cref{equ:alg-trivial:opt}.
\end{proof}
\section{Balanced Adaptive Greedy for \srma/}

In this section, we build on the simple algorithm NG to give a new combinatorial algorithm that obtains a logarithmic approximation.
As the proof of~\cref{thm:NG_Lower_bound} demonstrates, the shortfall of the previous algorithm is that it does not take into account the progress of agents in the process of covering functions. After the first step, the last agent $k$ has many more uncovered functions than other agents, but the algorithm does not prioritize the elements that can cover the functions of this agent. 
In other words, algorithm NG performs poorly in balancing agents' progress, which prevents the algorithm from getting rid of the polynomial dependence on $k$.
Based on this crucial observation, we propose the balanced adaptive greedy algorithm. 

We introduce some notation first.
Let $\rit$ be the set of functions of agent $a_i$ that are not satisfied earlier than time $t$.
Note that $\rit$ includes the functions that are satisfied exactly at time $t$.
Let $w(\rit)$ be the total weight of functions in $\rit$.
Recall that $W$ is the maximum total function weight among all agents.
Let $\cB=\set{(\frac{2}{3})^1\cdot W,(\frac{2}{3})^2\cdot W,\ldots}$ be a sequence, where each entry is referred to as the baseline in each iteration. 
%\ben{I am not sure what baseline means here}
Let $B_p$ be the $p$-th term of $\cB$, i.e., $B_p=(\frac{2}{3})^p\cdot W$.
Roughly speaking, in the $p$-th iteration, we want the uncovered functions of any agent to have a total weight at most $B_p$. 
Note that the size of $\cB$ is $\Theta(\log W)$.
Given a permutation $\pi$, let $\pi(t)$ be the element selected in the $t$-th time slot.
Let $\pi_t$ be the set of elements that are scheduled before or at time $t$, i.e., $\pi_t=\set{\pi(1),\ldots,\pi(t)}$.

\begin{algorithm}[tb]
\caption{Balanced Adaptive Greedy for \srma/}
\label{alg:comb}
\begin{algorithmic}[1]
\REQUIRE Ground elements $[n]$; Agent set $A$; A collection of $k$ weight vectors $\set{\vw^{(i)}\in\R^{m}_{+}|i\in[k]}$; A collection of $m\cdot k$ submodular functions $\set{\fij:2^{[n]}\to [0,1]|i\in[k],j\in[m]}$.
\ENSURE A permutation $\pi:[n]\to[n]$ of the elements.
\STATE $p\leftarrow 1$; $q\leftarrow 1$; $t\leftarrow 1$; $B_p\leftarrow (\frac{2}{3})^p\cdot W$
\STATE For each agent $a_i$, let $\rit$ be the set of functions of agent $a_i$ that are not satisfied earlier than time $t$.
\WHILE{there exists an agent $a_i$ with $w(\rit)>B_p$}
\label{line:outer:s}
\STATE $q\leftarrow 1$; $A_{p,q}\leftarrow \set{a_i\in A \mid w(\rit)>B_p}$.
% \STATE Let $A_{p,q}\subseteq A$ be all agents with $w(\rit)>B_p$.
\WHILE{$|A_{p,q}|\geq 0$}
\STATE $A'_{p,q}\leftarrow A_{p,q}$.
\label{line:drop_agents}
\WHILE{$|A_{p,q}|\geq \frac{3}{4}\cdot |A'_{p,q}|$}
\label{line:inner:s}
\STATE $F_{\pi_{t-1}}(e)\leftarrow \sum_{a_i\in A'_{p,q}}\sum_{j\in[m]}\wij\cdot \frac{\fij(\pi_{t-1}\cup\set{e})-\fij(\pi_{t-1})}{1-\fij(\pi_{t-1})},$ $\forall e\in U\setminus \pi_{t-1}$.
\label{line:greedy_1}
\STATE $\pi(t)\leftarrow \argmax_{e\in U\setminus \pi_{t-1}}F_{\pi_{t-1}}(e)$.
\label{line:greedy_2}
\STATE $t\leftarrow t+1$; $A_{p,q}\leftarrow \set{a_i\in A \mid w(\rit)>B_p}$.
\ENDWHILE
\label{line:inner:e}
\STATE $q\leftarrow q+1$; $A_{p,q}\leftarrow A_{p,q-1}$.
\ENDWHILE
\STATE $p\leftarrow p+1$.
\ENDWHILE
\label{line:outer:e}
\RETURN Permutation $\pi$.
\end{algorithmic}
\end{algorithm}

\cref{alg:comb} has two loops: outer iteration (line \ref{line:outer:s}-\ref{line:outer:e}) and inner iteration (line \ref{line:inner:s}-\ref{line:inner:e}). Let $p,q$ be the index of the outer and inner iterations, respectively. At the beginning of the $p$-th outer iteration, we remove all the satisfied functions and elements selected in the previous iterations, and we obtain a subinstance, denoted by $I_p$. At the end of the $p$-th outer iteration, \cref{alg:comb} ensures that the total weight of unsatisfied functions of each agent is at most $B_p=(\frac{2}{3})^p\cdot W$. Let $A_p$ be the set of agents whose total weight of unsatisfied functions is more than $B_p$. At the end of $(p,q)$-th inner iteration, \cref{alg:comb} guarantees that the number of agents with the total weight of unsatisfied function larger than $B_p$ is at most $(\frac{3}{4})^q\cdot |A_p|$.  Together, this implies there are at most $\Theta(\log W)$ outer iterations because $B_p$ decreases geometrically. Similarly, there are at most $\Theta(\log k)$ inner iterations for each outer iteration. 
 
Naturally, the algorithm chooses the element that gives the highest total weighted functional increase over all agents in $A_{p,q}$ and their corresponding submodular functions. When computing the marginal increase, each function is divided by how much is left to be fully covered. 
 
For technical reasons, during one iteration of line \ref{line:inner:s}-\ref{line:inner:e}, some agents may be satisfied, i.e., their total weight of unsatisfied functions is at most $B_p$.
Instead of dropping them immediately, we wait until $1/4$-the proportion of agents is satisfied, and then drop them together.
This is the role of line \ref{line:drop_agents} of \cref{alg:comb}.

%Let $p,q$ be the index of the outer and inner iterations, respectively.
%At the end of the $p$-th outer iteration, \cref{alg:comb} ensures that the total weight of unsatisfied functions of each agent is at most $B_p=\frac{2}{3}^p\cdot W$. Note that the total weight of functions of each agent that are not satisfied decreases exponentially in $p$ and the number of outer iterations can be bounded by $O(\log W)$.
%At the beginning of the $p$-th outer iteration, we remove all satisfied functions and selected element in the last iteration, and we obtain a subinstance, denoted by $I_p$.

%The inner iterations will ensure that \emph{all} agents are satisfied to  at least $(1-(\frac{2}{3})^p)\cdot W$ of their total weights.

%Let $A_p$ be the set of agents whose total weight of unsatisfied functions are more than $B_p$.  \cref{alg:comb} uses $O(\log k)$ inner iterations.
%At the end of $(p,q)$-th iteration, \cref{alg:comb} guarantees that the number of agents with the total weight of unsatisfied function lager than $B_p$ is at most $(1-(\frac{1}{4})^q)\cdot |A_p|$.
%Note that, during one iteration of line \ref{line:inner:s}-\ref{line:inner:e}, some agents may be satisfied, i.e., their total weight of unsatisfied functions is at most $B_p$.
%Instead of dropping them immediately, we wait until $\frac{1}{4}$-proportion of agents are satisfied, and then we drop them together.
%This is the role of line  \ref{line:drop_agents} of \cref{alg:comb}. \ben{Should we describe in works which function we pick}
\section{Analysis}

Given an arbitrary instance $I$ of \srma/, let $\opt(I)$ be an optimal weighted cover time of instance $I$.
Our goal is to show \cref{thm:main}. 
Recall that $W=\max_{i\in[k]}\sum_{j=1}^{m}\wij$ is the maximum total weight among all agents; we assume that all weights are integers. We will first show the upper bound depending on $\log W$. We will later show how to obtain another bound depending on $\log n$ at the end of the analysis, where $n$ is the number of elements.

\begin{theorem}
\label{thm:main}
%There exists an algorithm for \srma/ with an approximation ratio of 
\cref{alg:comb} obtains an approximation ratio of $O( \ln(1 / \epsilon)\log(\min\{n,W\}) \log k )$, where $\epsilon$ is the minimum non-zero marginal value among all $m\cdot k$ monotone submodular functions, $k$ is the number of agents and $W$ is the maximum total integer weight of all functions among all agents.
\end{theorem}

We first show a lower bound of $\opt(I)$.
Let $\pi^*$ be an optimal permutation of the ground elements.
Let $\orit$ be the set of functions of agent $a_i$ that are not satisfied earlier than time $t$ in $\pi^*$.
Note that $\orit$ includes the functions that are satisfied exactly at time $t$.
Let $w(\orit)$ be the total weight of functions in $\orit$.
Let $T^*$ be the set of times from $t=1$ to the first time that all agents in the optimal solution satisfy $w(\orit)\leq\alpha\cdot W$, where $\alpha\in(0,1)$ is a constant.
Let $E(T^*)$ be the corresponding elements in the optimal solution $\pi^*$, i.e., $E(T^*)=\set{\pi^*(1),\ldots,\pi^*(\abs{T^*})}$.

\begin{figure}[tb]
    \centering
    \includegraphics[width=8cm]{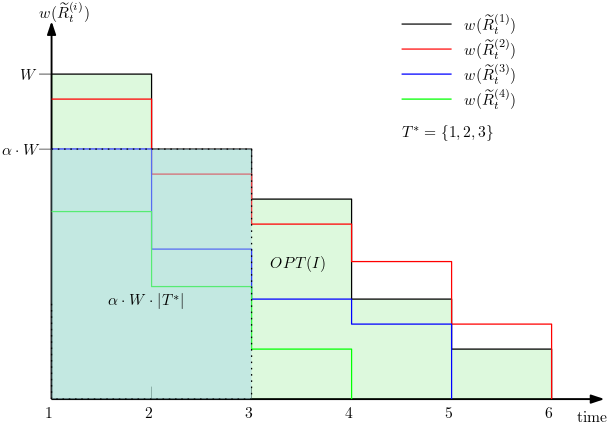}
    \caption{An illustration of the optimal value's lower bound. There are four agents in the figure. The area enclosed by the $x$-axis, $y$-axis, and curve $w(\orit)$ is the total weighted cover time of agent $a_i$. The optimal value will be the largest one. Then, the total area of $\alpha\cdot W\cdot \abs{T^*}$ will be completely included in the area that is formed by the optimal solution.}
    \label{fig:lower_bound}
\end{figure}

We have the following lower bound of the optimal solution.
An example of \cref{fac:lower_bound} can be found in \cref{fig:lower_bound}.

\begin{fact}{(Lower Bound of $\opt$)}
$\alpha\cdot W\cdot \abs{T^*} \leq \opt(I)$.
\label{fac:lower_bound}
\end{fact}

% See \cref{fig:lower_bound}.

Recall that $I_p$ is the subinstance of the initial instance at the beginning of the $p$-th iteration of \cref{alg:comb}.
Therefore, we have $\opt(I_p)\leq\opt(I)$ for all $p\leq\ceil{\log W}$.
Let $\pi^*(I_p)$ be the optimal permutation of instance $I_p$.
Recall that $B_p=(\frac{2}{3})^p\cdot W$.
Let $T_p^*$ be the set of times from $t=1$ to the first time such that all agents satisfy $w(\orit)\leq\frac{1}{12}\cdot B_p$ in $\pi^*(I_p)$.
By \cref{fac:lower_bound}, we know that $\frac{1}{12}\cdot B_p\cdot \abs{T^*_p}\leq \opt(I_p) \leq \opt(I)$.
Let $T_p$ be the set of times that are used by \cref{alg:comb} in the $p$-th outer iteration, i.e., $T_p = {t_{p}, \ldots, t_{p+1} - 1}$ where $t_p$ is the time at the beginning of the $p$-th outer iteration.
Let $E(T_p)$ be the set of elements that are selected by \cref{alg:comb} in the $p$-th outer iteration.
Note that $\abs{T_p}$ is the length of the $p$-th outer iteration.
Recall that every outer iteration contains $\Theta(\log k)$ inner iterations.
In the remainder of this paper, we denote the $q$-th inner iteration of the $p$-th outer iteration as $(p,q)$-th iteration.

We now give a simple upper bound of the solution returned by the algorithm.

\begin{fact}{(Upper Bound of $\alg$)}
$\alg(I)\leq \sum_{p} \abs{T_p}\cdot B_{p-1}=\sum_{p}\frac{2}{3}\cdot \abs{T_p} \cdot B_p$.
\label{fac:upper_bound}
\end{fact}

Our analysis relies on the following key lemma (\cref{lem:key}).
Roughly speaking, \cref{lem:key} measures how far the algorithm is behind the optimal solution after each outer iteration.

\begin{lemma}
For any $p\leq\ceil{\log W}$, $\abs{T_p}\leq O\left((1+\ln(\frac{1}{\epsilon}))\cdot \log k\right)\cdot \abs{T_p^*}$, where $\epsilon$ is the minimum non-zero marginal value and $k$ is the number of agents.
\label{lem:key}
\end{lemma}

%Based on \cref{lem:key}, we now prove \cref{thm:main}.
In the following, we first show that if \cref{lem:key} holds, \cref{thm:main} can be proved easily, and then give the proof of \cref{lem:key}.

\begin{proof}[Proof of \cref{thm:main}]
Combining \cref{fac:lower_bound}, \cref{fac:upper_bound} and \cref{lem:key}, we have
$\alg(I)\leq O((1+\ln(\frac{1}{\epsilon})) \log k\log W)\cdot\opt(I)$. 
Formally, we have the following inequalities:
\begin{align*}
\alg(I)&\leq\sum_{p}\frac{2}{3}\cdot \abs{T_p} \cdot B_p \tag*{[Due to \cref{fac:upper_bound}]} \\
&\leq\sum_{p}O((1+\ln(\frac{1}{\epsilon}))\cdot \log k) \cdot \abs{T_p^*} \cdot B_p \tag*{[Due to \cref{lem:key}]} \\
&\leq\sum_{p}O((1+\ln(\frac{1}{\epsilon}))\cdot \log k) \cdot \opt(I_p) \tag*{[Due to \cref{fac:lower_bound}]} \\
&\leq\sum_{p}O((1+\ln(\frac{1}{\epsilon}))\cdot \log k) \cdot \opt(I) \tag*{[Due to $\opt(I_b)\leq\opt(I)$]} \\
&\leq O\left((1+\ln(\frac{1}{\epsilon}))\cdot \log k \cdot \log W\right) \cdot \opt(I) \tag*{[Due to $p\leq \ceil{\log W}$]}
\end{align*}
To obtain the other bound of $\log n$ claimed in \cref{thm:main}, we make a simple observation. 
Once the total weight of the uncovered functions drops below $W / n$ for any agent, they incur at most $(W / n) n = W$ cost in total afterward as there are at most $n$ elements to be ordered. Until this moment, $O(\log n)$ different values of $p$ were considered. Thus, in the analysis, we only need to consider $O(\log n)$ different values of $p$, not $O(\log W)$. This gives the desired approximation guarantee. 
\end{proof}

%It now remains to show \cref{lem:key}.

Now we prove \cref{lem:key}. Let $T_{p,q}$ be the times that are used by \cref{alg:comb} in the $(p,q)$-th iteration.
Let $E(T_{p,q})$ be the set of elements that are selected by \cref{alg:comb} in the $(p,q)$-th iteration.
Note that $\abs{T_{p,q}}$ is the length of the $q$-th inner iteration in $p$-th outer inner iteration.
Let $T^*_{p,q}$ be the set of times from $t=1$ to the first time such that all agents in $A_{p,q}$ satisfy $w(\orit)\leq \frac{1}{12}\cdot B_p$ in $\pi^*(I_p)$. %The correctness of \cref{lem:key} relies on the following lemma.
Observe that if, for any $p\leq\ceil{\log W}$ and $q\leq \ceil{\log k}$, we have: %\cref{lem:key_base} holds

% \begin{lemma}
% For any $p\leq\ceil{\log W}$ and $q\leq \ceil{\log k}$, we have $\abs{T_{p,q}}\leq \left(1+\ln(\frac{1}{\epsilon})\right)\cdot \abs{T_{p,q}^*}$.
% \label{lem:key_base}
% \end{lemma}

\begin{equation}
\abs{T_{p,q}}\leq \left(1+\ln(\frac{1}{\epsilon})\right)\cdot \abs{T_{p,q}^*},
\label{lem:key_base}
\end{equation}
%Based on \cref{lem:key_base}, 
the proof of \cref{lem:key} is straightforward because $T_{p,q}^*\subseteq T_{p}^*$ for any $p\leq\ceil{\log W}$ and the number of inner iterations is $\Theta(\log k)$.
%Formally, we have the following inequalities:
Thus, the major technical difficulty of our algorithm is to prove \cref{lem:key_base}.
Let $\pi$ be the permutation returned by \cref{alg:comb}.
Note that the time set $[n]$ can be partitioned into $O(\log W\cdot \log k)$ sets.
Recall that $T_{p,q}$ is the time set that are used in the $(p,q)$-th iteration and $A_{p,q}$ is the set of unsatisfied agents at the beginning of the $(p,q)$-th iteration, i.e., $A_{p,q}=\set{a_i\in A\mid w(\rit)>B_p}$. 
For notation convenience, we define $F_{\pi_{t-1}}(e)$ as follows:
% \begin{align*}
% & F_{\pi_{t-1}}(e) \\
% &:=\sum_{a_i\in A_{p,q}}\sum_{j\in[m]}\wij\cdot\frac{\fij(\pi_{t-1}\cup\set{e})-\fij(\pi_{t-1})}{1-\fij(\pi_{t-1})}
% \end{align*}
$$
F_{\pi_{t-1}}(e):=\sum_{a_i\in A_{p,q}}\sum_{j\in[m]}\wij\cdot\frac{\fij(\pi_{t-1}\cup\set{e})-\fij(\pi_{t-1})}{1-\fij(\pi_{t-1})}
$$

We present two technical lemmas sufficient to prove \cref{lem:key_base}.

\begin{lemma}
For any $p\leq\ceil{\log W}$ and $q\leq \ceil{\log k}$, we have 
$$
\sum_{t\in T_{p,q}} F_{\pi_{t-1}}(e_t) \leq \left(1+\ln(\frac{1}{\epsilon})\right)\cdot \abs{A_{p,q}} \cdot B_{p-1}
$$
\label{lem:key_base_upper}
\end{lemma}

\begin{lemma}
For any $p\leq\ceil{\log W}$ and $q\leq \ceil{\log k}$, we have 
$$
\sum_{t\in T_{p,q}} F_{\pi_{t-1}}(e_t) \geq\frac{2}{3} \cdot \frac{\abs{T_{p,q}}}{\abs{T_{p,q}^*}}\cdot \abs{A_{p,q}} \cdot B_p
$$
\label{lem:key_base_lower}
\end{lemma}

The proof of~\cref{lem:key_base_upper} uses a property of a monotone function (\cref{pro:mono}). Note that the proof of \cref{pro:mono} can be found in \cite{DBLP:conf/soda/AzarG11} and \cite{DBLP:journals/talg/ImNZ16}.
Here we only present the statement for completeness. Proving \cref{lem:key_base_lower} uses a property that is more algorithm-specific (\cref{pro:greedy}). 
Recall that in each step $t$ of an iteration $(p,q)$, \cref{alg:comb} will greedily choose the element $e_t$ among all unselected elements such that $F_{\pi_{t-1}}(e_t)$ is maximized. Then we have that the inequality $F_{\pi_{t-1}}(e_t) \geq F_{\pi_{t-1}}(e)$ holds for any element $e\in U$, and hence, for any element selected by the optimal solution. By an average argument, we can build the relationship between $F_{\pi_{t-1}}(e)$ and $|T_{p,q}^*|$, and prove \cref{lem:key_base_lower}. 
\cref{lem:key_base} follows from \cref{lem:key_base_upper} and \cref{lem:key_base_lower} simply by concatenating the two inequalities.

%Now, we are ready to prove \cref{lem:key_base_upper} and \cref{lem:key_base_lower}.
% The correctness of \cref{lem:key_base_upper} relies on a property of monotone function stated in \cref{pro:mono}.

\begin{proposition}{(Claim 2.4 in \cite{DBLP:journals/talg/ImNZ16})}
Given an arbitrary monotone function $f:2^{[n]}\to[0,1]$ with $f([n])=1$ and sets $\emptyset=S_0\subseteq S_1 \subseteq \ldots \subseteq S_h \subseteq [n]$, we have
$$
\sum_{i=1}^{h}\frac{f(S_i)-f(S_{i-1})}{1-f(S_{i-1})} \leq 1+\ln{\frac{1}{\delta}},
$$
where $\delta>0$ is such that for any $A\subseteq B$, if $f(B)-f(A)\geq 0\implies f(B)-f(A)\geq\delta$.
\label{pro:mono}
\end{proposition}

\begin{proof}
[Proof of \cref{lem:key_base_upper}]
Consider an arbitrary pair of $p\leq\ceil{\log W}$ and $q\leq\ceil{\log k}$.
Recall that $A_{p,q}$ is the set of unsatisfied agents, i.e., for any $a_i\in A_{p,q}$, we have $w(\rit)>B_p$ at the beginning of the $(p,q)$-th iteration.
Recall that $T_{p,q}$ is a set of times that are used in the iteration $(p,q)$.
Let $\bar{t}$ be the last time in $T_{p,q-1}$ if $q\geq 2$; otherwise let $\bar{t}$ be the first time of $T_{p,q}$.
Note that, during the whole $q$-th inner iteration, the agent set that we look at keeps the same (line \ref{line:drop_agents} of \cref{alg:comb}).
Then, we have
\begin{align*}
&\sum_{t\in T_{p,q}} F_{\pi_{t-1}}(e_t) \\
&=\sum_{t\in T_{p,q}}\sum_{a_i\in A_{p,q}}\sum_{j\in\rit}\wij\cdot\frac{\fij(\pi_{t-1}\cup\set{e_t})-\fij(\pi_{t-1})}{1-\fij(\pi_{t-1})} \\
&\leq \sum_{a_i\in A_{p,q}}\sum_{j\in\ritb}\sum_{t\in T_{p,q}}\wij\cdot\frac{\fij(\pi_{t-1}\cup\set{e_t})-\fij(\pi_{t-1})}{1-\fij(\pi_{t-1})} \tag*{[Due to $\rit\subseteq \ritb$]}\\
&= \sum_{a_i\in A_{p,q}}\sum_{j\in\ritb}\wij\cdot\sum_{t\in T_{p,q}}\frac{\fij(\pi_{t-1}\cup\set{e_t})-\fij(\pi_{t-1})}{1-\fij(\pi_{t-1})} \\
&\leq \left( 1+\ln\frac{1}{\epsilon} \right)\cdot \sum_{a_i\in A_{p,q}}\sum_{j\in\ritb} \wij \tag*{[Due to \cref{pro:mono}]} \\
&\leq\left( 1+\ln\frac{1}{\epsilon} \right) \cdot \sum_{a_i\in A_{p,q}} B_{p-1} \tag*{[Due to the definition of $\bar{t}$]} \\
&\leq \left( 1+\ln\frac{1}{\epsilon} \right) \cdot \abs{A_{p,q}} \cdot B_{p-1}
\end{align*}
\end{proof}

The last piece is proving \cref{lem:key_base_lower}, which relies on following property of the greedy algorithm.

\begin{proposition}
For any $p\leq\ceil{\log W}$, $q\leq\ceil{\log k}$, consider an arbitrary time $t\in T_{p,q}$ and let $e_t$ be the element that is selected by \cref{alg:comb} at time $t$.
For any $e\in E(T_{p,q}^*)$, we have $F_{\pi_{t-1}}(e_t)\geq F_{\pi_{t-1}}(e)$.
\label{pro:greedy}
\end{proposition}

\begin{proof}
[Proof of \cref{lem:key_base_lower}]
By \cref{pro:greedy}, we know that $F_{\pi_{t-1}}(e_t)\geq F_{\pi_{t-1}}(e)$ for any $e\in E(T_{p,q}^*)$.
Thus, by average argument, we have \cref{equ:greedy} for any $t\in T_{p,q}$.
Recall that $e_t$ is the element that is selected by \cref{alg:comb} at time $t$.
\begin{equation}
F_{\pi_{t-1}}(e_t)\geq \frac{1}{\abs{E(T_{p,q}^*)}}\cdot \sum_{e\in E(T_{p,q}^*)} F_{\pi_{t-1}}(e) 
\label{equ:greedy}    
\end{equation}
Let $t^*$ be the last time in $T_{p,q}^*$, i.e., $t^*$ is the first time such that all agents in $A_{p,q}$ satisfy $w(\orit)\leq\frac{1}{12}\cdot B_p$ in $\pi^*(I_p)$, where $\pi^*(I_p)$ is the optimal permutation to the subinstance $I_p$.
Let $\widetilde{t}$ be the last time in $T_{p,q}$. 
Note that, for any $t\in T_{p,q}$, we know that there are at most $\frac{1}{4}\cdot\abs{A_{p,q}}$ agents satisfying $w(\rit)\leq B_p$.
Thus, we have
\begin{equation}
\sum_{a_i\in A_{p,q}}w(\rit) > \frac{3}{4} \cdot \abs{A_{p,q}} \cdot B_p, \forall t\in T_{p,q} 
\label{equ:proportion}
\end{equation}
Since $\widetilde{t}\in T_{p,q}$, we know that $\sum_{a_i\in A_{p,q}}w(\ritw)>\frac{3}{4}\cdot\abs{A_{p,q}}\cdot B_p$.
Thus, we have the following inequalities.
\begin{align*}
&\sum_{t\in T_{p,q}}F_{\pi_{t-1}}(e_t) \\
&=\sum_{t\in T_{p,q}}\sum_{a_i\in A_{p,q}}\sum_{j\in\rit}\wij\cdot\frac{\fij(\pi_{t-1}\cup\set{e_t})-\fij(\pi_{t-1})}{1-\fij(\pi_{t-1})} \\
&\geq\sum_{t\in T_{p,q}}\frac{1}{\abs{E(T^*_{p,q})}}\sum_{e\in E(T_{p,q}^*)}\sum_{a_i\in A_{p,q}}\sum_{j\in\rit}\wij\cdot\frac{\fij(\pi_{t-1}\cup\set{e})-\fij(\pi_{t-1})}{1-\fij(\pi_{t-1})} \tag*{[Due to \cref{equ:greedy}]} \\
&\geq\sum_{t\in T_{p,q}}\frac{1}{\abs{E(T^*_{p,q})}}\sum_{a_i\in A_{p,q}}\sum_{j\in\rit}\wij\cdot\frac{\fij(\pi_{t-1}\cup E(T_{p,q}^*))-\fij(\pi_{t-1})}{1-\fij(\pi_{t-1})} \tag*{[Due to $\fij$ is submodular]} \\
&\geq\sum_{t\in T_{p,q}}\frac{1}{\abs{E(T_{p,q})^*}}\cdot\sum_{a_i\in A_{p,q}}\sum_{j\in\ritw\setminus\orits}
\wij\cdot\frac{\fij(\pi_{t-1}\cup E(T_{p,q}^*))-\fij(\pi_{t-1})}{1-\fij(\pi_{t-1})} \tag*{[Due to $\ritw\setminus\orits \subseteq \ritw$]} \\
&\geq\sum_{t\in T_{p,q}}\frac{1}{\abs{E(T_{p,q}^*)}}\cdot \sum_{a_i\in A_{p,q}} w(\ritw\setminus\orits) \tag*{[Due to $\fij(E(T^*_{p,q}))=1$ for any $j\notin \orits$]} \\
&\geq\sum_{t\in T_{p,q}}\frac{1}{\abs{E(T_{p,q}^*)}}\cdot \sum_{a_i\in A_{p,q}}\left( w(\ritw)-w(\orits) \right) \\
&\geq\frac{\abs{T_{p,q}}}{\abs{T_{p,q}^*}}\cdot \left( \sum_{a_i\in A_{p,q}} w(\ritw)-\frac{1}{12}\cdot \abs{A_{p,q}}\cdot B_p \right) \tag*{[Due to the definition of $t^*$]} \\
&\geq\frac{\abs{T_{p,q}}}{\abs{T_{p,q}^*}}\cdot \left( \frac{3}{4}\cdot\abs{A_{p,q}}\cdot B_p-\frac{1}{12}\cdot \abs{A_{p,q}}\cdot B_p \right) \tag*{[Due to \cref{equ:proportion}]}\\
&=\frac{2}{3}\cdot\frac{\abs{T_{p,q}}}{\abs{T^*_{p,q}}}\cdot\abs{A_{p,q}}\cdot B_p
\end{align*}
\end{proof}

\begin{proof}
[Proof of \cref{lem:key}]
Combining the above lemmas, we have
\begin{align*}
\abs{T_p} &= \sum_{q\leq\ceil{\log k}} \abs{T_{p,q}} \\
&\leq\left( 1+\ln(\frac{1}{\epsilon}) \right)\cdot \ceil{\log k} \cdot \abs{T_{p,q}^*} \tag*{[Due to \cref{lem:key_base_upper} and \cref{lem:key_base_lower}]} \\
&\leq O\left( (1+\ln(\frac{1}{\epsilon})) \cdot \log k \right) \cdot \abs{T_{p}^*}  \tag*{[Due to $\abs{T_{p,q}^*}\leq \abs{T_{p}^*}$]} 
\end{align*}
\end{proof}

\iffalse
$T_{p,q}^*$ is the set of times from $t=1$ to the first time such that all agents in $A_{p,q}$ satisfy $w(\orit)\leq\frac{1}{12}\cdot B_p$ in $\pi^*(I_p)$, where $\pi^*(I_p)$ is the optimal permutation to the subinstance $I_p$.
Let $E(T^*_{p,q})$ be the set of elements that are selected in $\pi^*(I_p)$ in times $T^*_{p,q}$. 
We need to use the property of \cref{alg:comb} which is stated in \cref{pro:greedy}.
\fi

%The correctness of \cref{lem:key_base} relies on the following two lemmas (\cref{lem:key_base_upper} and \cref{lem:key_base_lower}).
%Combining these two lemmas, \cref{lem:key_base} directly follows.
%In the following, we first show \cref{lem:key_base_upper}, which replies on a property of a monotone function.
%More details can be found in \cref{app:balance_greed}.

\iffalse
stated in \cref{pro:mono}.
The proof of \cref{pro:mono} can be found in \cite{DBLP:conf/soda/AzarG11} and \cite{DBLP:journals/talg/ImNZ16}.
Here we only present the statement for completeness.
\fi

\iffalse
\begin{proposition}{(Claim 2.4 in \cite{DBLP:journals/talg/ImNZ16})}
Given an arbitrary monotone function $f:2^{[n]}\to[0,1]$ with $f([n])=1$ and sets $\emptyset=S_0\subseteq S_1 \subseteq \ldots \subseteq S_h \subseteq [n]$, we have
$$
\sum_{i=1}^{h}\frac{f(S_i)-f(S_{i-1})}{1-f(S_{i-1})} \leq 1+\ln{\frac{1}{\delta}},
$$
where $\delta>0$ is such that for any $A\subseteq B$, if $f(B)-f(A)\geq 0\implies f(B)-f(A)\geq\delta$.
\label{pro:mono}
\end{proposition}
\fi

%\smallskip

\section{Inapproximability Results for SRMA and Tight Approximation for GMSC}
\label{sec:mssc+hardness}

In this section, we consider a special case of \srma/, which is called Generalized Min-sum Set Cover for Multiple Agents (GMSC), and leverage it to give a lower bound on the approximation ratio of \srma/. We first state the formal definition.
%The formal definition of GMSC can be found in the second subsection of the current section.

\paragraph{GMSC for Multiple Agents.} An instance of this problem consists of a ground set $U:=[n]$, and a set of $k$ agents $A:=\set{a_1,\ldots,a_k}$.
Every agent $a_i$, $i\in[k]$, has a collection $\cS^{(i)}$ of $m$ subsets defined over $U$, i.e., $\cS^{(i)}=\set{S_1^{(i)},\ldots,S_m^{(i)}}$ with $S_j\subseteq U$ for all $j\in[m]$. 
Each set $\sij$ comes with a coverage requirement $K(\sij)\in\set{1,2,\ldots,\abs{\sij}}$.
Given a permutation $\pi:[n]\to[n]$ of the elements in $U$, the cover time $\cov(\sij,\pi)$ of set $\sij$ is defined to the first time such that $K(\sij)$ elements from $\sij$ are hit in $\pi$.
The goal is to find a permutation $\pi$ of the ground elements such that the maximum total cover time is minimized among all agents, i.e., finding a permutation $\pi$ such that $\max_{i\in[k]}\left\{ \sum_{j=1}^{m} \cov(\sij,\pi) \right\}$ is minimized.

\subsection{Inapproximability Results}

We first show a lower bound of GMSC.
The basic idea is to show that GMSC captures the classical set cover instance as a special case. To see the reduction, think of each element $e$ in the set cover instance as an agent in the instance of GMSC and each set $S$ as a ground element.
Every agent has only one set, and all of the coverage requirements are $1$.
Then, a feasible solution to the set cover instance will be a set of subsets (elements in GMSC instance) that covers all elements (satisfies all agents in GMSC instance).
Thus, the optimal solutions for these two instances are equivalent.

\begin{lemma}
For the generalized min-sum set cover for multiple agents problem, given any constant $c<1$, there is no $c\cdot \ln k$-approximation algorithm unless P=NP, where $k$ is the number of agents.
\label{thm:hardness:mssc}
\end{lemma}

\begin{proof}
To prove the claim, we present a approximation-preserving reduction from the set cover problem.
Given an arbitrary instance of set cover $(E,\cC)$, where $E=\set{e_1,\ldots,e_p}$ is the set of ground elements and $\cC=\set{C_1,\ldots,C_q}$ is a collection of subsets defined over $E$.
By \cite{DBLP:journals/jacm/Feige98}, we know that it is NP-hard to approximate the set cover problem within $c\cdot\ln p$ factor for any  constant $c<1$, where $p$ is the number of ground set elements.

For any element $e_i\in E$, let $V(e_i)$ be the collection of the subsets that contain $e_i$, i.e., $V(e_i)=\set{C_j\in \cC \mid e_i\in C_j}$.
For each element $e_i\in E$, we create an agent $a_i$ in our problem.
For each subset $C_j\in\cC$ of in the set cover instance, we create a ground set element $j\in U$.
In total, there are $p$ agents and $q$ ground set elements, i.e., $\abs{A}=p$ and $\abs{U}=q$.
Every element $e_i$ in the set cover instance has an adjacent set $V(e_i)$ which is a collection of the subsets.
Every subset in the set cover instance corresponds an element in our problem.
Thus, $V(e_i)$ corresponds a subset of $U$ of our problem.
Let $S^{(i)}$ be the corresponding subsets of $V(e_i)$ in our problem ($S^{(i)}\subseteq U$).
Every agent $a_i$ has only one set $S^{(i)}$.
And all sets come with a same covering requirement $1$.

In such an instance of our problem, the total cover time of each agent is exactly the same as the hitting time of of her only set.
Thus, the maximum total cover time among all agents is exactly the same as the length of element permutation.
Therefore, the optimal solution to the set cover instance and the constructed instance of our problem can be easily converted to each other.
Moreover, both the optimal solutions share a same value which completes the proof.
\end{proof}

Note that our problem admits the submodular ranking problem as a special case which is $c\ln(1 / \epsilon)$-hard to approximate for some constant $c>0$ \cite{DBLP:conf/soda/AzarG11}.
Thus, our problem has a natural lower bound $\Omega(\ln(1 / \epsilon))$.
Hence, by combining \cref{thm:hardness:mssc} and the lower bound of classical submodular ranking, one can easily get a $\Omega(\ln(1 / \epsilon)+\log k)$ lower bound for \srma/ .

\begin{theorem}
The problem of min-max submodular ranking for multiple agents cannot be approximated within a factor of $c\cdot\left( \ln(1 / \epsilon)+\log k \right)$ for some constant $c>0$ unless P$=$NP, where $\epsilon$ is the minimum non-zero marginal value and $k$ is the number of agents.
\label{cor:hardness}
\end{theorem}

\subsection{Tight Approximation for GMSC}

This subsection mainly shows the following theorem.

\begin{theorem}
There is a randomized algorithm that achieves $\Theta(\log k)$-approximation for generalized
min-sum set cover for multiple agents, where $k$ is the number of agents.
\label{thm:mssc}
\end{theorem}

%Now we show a $O(\log k)$-approximation algorithm for the generalized min-sum set cover for multiple agents (\cref{thm:mssc}).
Our algorithm is LP-based randomized algorithm, which mainly follows from \cite{DBLP:conf/soda/BansalGK10}.
In the LP relaxation, the variable $x_{e,t}$ indicates for whether element $e\in U$ is scheduled at time $t\in[n]$.
The variable $y_{\sij, t}$ indicates for whether set $\sij$ has been covered before time $t\in[n]$. 
To reduce the integrality gap, we introduce an exponential number of constraints. 
We show that the ellipsoid algorithm can solve such an LP.
We partition the whole timeline into multiple phases based on the optimal fractional solution.
In each phase, we round variable $x_{e,t}$ to obtain an integer solution. 
The main difference from \cite{DBLP:conf/soda/BansalGK10}, we repeat the rounding algorithm $\Theta(\log k)$ time and interleave the resulting solutions over time to avoid the bad events for all $k$ agents simultaneously.
Then, we can write down the following LP.
Our objective function is min max which is not linear.
We convert the program into linear by using the standard binary search technique.
Namely, fix an objective value $T$ and test whether the linear program has a feasible solution.

\begin{align}
    % \min T & \nonumber\\
    \sum_{t \in [n]} \sum_{\sij \in \cS^{(i)}} 1 - y_{\sij, t} &\leq T \quad \forall a_i \in A \nonumber\\
    \sum_{e \in U} x_{e, t} &= 1 \quad \forall t \in [n] \nonumber\\
    \sum_{t \in [n]} x_{e, t} &= 1 \quad \forall e \in U \nonumber\\
    \sum_{e \in \sij \setminus B} \sum_{t' < t} x_{e, t'} &\geq (K(\sij) - |B|) y_{\sij, t} \quad \forall \sij \in \cS^{(i)}, B \subseteq \sij, t \in [n] \label{eqn: g-knap}\\
    x_{e, t}, y_{\sij, t} &\in [0, 1]  \nonumber
\end{align}

Note that the size of the LP defined above is exponential.
By the same proof in~\cite{DBLP:conf/soda/BansalGK10}, we have the following lemma, and thus, the LP can be solved polynomially.

\begin{lemma}
There is a polynomial time separation oracle for the LP defined above.
\end{lemma}

Now we describe our algorithm. 
Firstly, we solve the LP above and let $(x^*, y^*, T^*)$ be the optimal fractional solution of the LP. 
Then, our rounding algorithm consists of $\ceil{\log n}$ phases. 
Consider one phase $\ell \leq \ceil{\log n}$. 
We first independently run $2 \log k$ times \cref{alg:rounding}. 
Let $\pi_\ell^q$ be the $q$-th rounding solution for phase $\ell$. 
After $\lceil \log n \rceil$ phases, we outputs $\pi = \pi_1^1 \ldots \pi_1^{\log k} \ldots \pi_{\lceil \log n \rceil}^1 \ldots \pi_{\lceil \log n \rceil}^{\log k}$. 
Note that the concatenation of $\pi_\ell^q$ only keeps the first occurrence of each element.

\begin{algorithm}[htp]
\caption{LP Rounding for phase $\ell$}
\label{alg:rounding}
\begin{algorithmic}[1]
    \STATE let $t_\ell = 2^\ell$.
    \STATE let $\bar{x}_{e, \ell} \gets \sum_{t' < t_\ell}x_{e, t'}^*$.
    \STATE let $p_{e, \ell} = \min \{ 1, 8 \cdot \bar{x}_{e, \ell}\}$. Pick $e \in U$ with probability $p_{e, \ell}$.
    \STATE let $\pi_\ell$ be an arbitrary order of picked elements.
    \STATE if $|\pi_\ell| > 16 \cdot 2^\ell$, then let $\pi$ be empty.
\end{algorithmic}
\end{algorithm}

For any set $\sij$, let $t_{\sij}^*$ be the last time $t$ such that $y_{\sij, t} \leq \frac{1}{2}$. 
Since for time $t \leq t^*_{\sij}$, $1 - y_{\sij, t} \geq \frac{1}{2}$, the following \cref{lem:g-low} is immediately proved.
This gives us a lower bound of the optimal solution.

\begin{lemma}
\label{lem:g-low}
For any $i\in[k]$ and $j\in[m]$, $T^* \geq \frac{1}{2} \sum_{\sij \in \cS^{(i)}} t^*_{\sij}$.
\end{lemma}

% The correctness of \cref{alg:rounding} relies on following \cref{lem:mssc:ratio}.
% This states the event that total cover time of all agents is at most $O(\log k)$ times optimal fractional solution happens in a high probability.

% \begin{lemma}
% The probability that the total cover time is at most $O(\log k) \cdot T^*$ for all agents is at least $1-\frac{1}{k}$, i.e., $\Pr[\max_{i \in [k]} \sum_{j=1}^m \cov(\sij, \pi) \leq 1024 \log k \cdot T^*] \geq 1 - \frac{1}{k}$, where $k$ is the number of agents.
% \label{lem:mssc:ratio}
% \end{lemma}

% To prove \cref{lem:mssc:ratio}, we need to show that the probability that the total cover time of each agent is at $1-\frac{1}{k^2}$.
% Then, \cref{lem:mssc:ratio} can be proved by standard union bound.
% To show the probability of a single agent, we need the help of \cref{lem:ori-approx} which stats the excepted cover time of any agent is at most some constant factor of the optimal fraction value in arbitrary independent running of \cref{alg:rounding}.

% \begin{lemma}
%     \label{lem:ori-approx}
%     For any $q \leq 2\ceil{\log k}$, the expected cover time of any agent given by $\pi^q = \pi^q_1 \ldots \pi^q_{\ceil{\log n}}$ is at most $512 \cdot T^*$.
% \end{lemma}

% Then, the probability that the total cover of an arbitrary agent less  than the desired bound can be proved by Markov's inequality.

Using the same techniques in~\cite{DBLP:conf/soda/BansalGK10}, we show the following three lemmas.
\begin{lemma}
    \label{lem:r-Knap}
    For any agent $a_i$ and set $\sij$ in phase $\ell$ such that $t^*_{\sij} \leq t_\ell$, the probability that $K(\sij)$ elements from $\sij$ are not picked in phase $\ell$ is at most $e^{-9/8}$.
\end{lemma}
\begin{proof}
    Consider set $\sij$. Let $S_g = \set{e \in \sij : \bar{x}_{e, \ell} \geq 1/8}$. For $e \in S_g$, by Step 3 of \Cref{alg:rounding}, we know that element $e$ is picked in phase $\ell$. Therefore, if $|S_g| \geq K(\sij)$, the lemma holds.
    
    Thus, we only need to focus on $|S_g| < K(\sij)$. Since $t^*_{\sij} \leq t_\ell$, we know $y^*_{\sij, t_\ell} > 1/2$. Plugging $B = S_g$ into \Cref{eqn: g-knap}, we have
    \begin{align*}
        \sum_{e \in \sij \setminus S_g} \bar{x}_{e, \ell} &= \sum_{e \in \sij \setminus S_g} \sum_{t' < t_\ell} x^*_{e, t'} \\
        &\geq  (K(\sij) - |S_g|) y^*_{\sij, t_\ell}\\
        &\geq 1/2 (K(\sij) - |S_g|) 
    \end{align*}
    Therefore, the expected number of elements from $\sij \setminus S_g$ picked in phase $\ell$ is 
    $$ \sum_{e \in \sij \setminus S_g} 8 \bar{x}_{e,\ell} \geq 4 (K(\sij) - |S_g|)$$
    
    Since every element is independently picked, we can use the following Chernoff bound: 
    If $X_1, ..., X_n$ are independent {0,1}-valued random variables with $X = \sum_{i=1}^n X_i$ such that $\mu = \E[X]$, 
    then $ \Pr[X < (1-\beta) \mu] \leq \exp(-\frac{\beta^2}{2} \mu) $. 
    Since the expected number of elements picked is at least $4 (K(\sij) - |S_g|) \geq 4$, set $\beta = 3/4$ and $\mu = 4$, the probability the less than $K(\sij) - |S_g|$ were picked is at most $e^{-9/8}$. Combining with the event that all elements in $S_g$ were picked, \Cref{lem:r-Knap} follows.

\end{proof}

\begin{lemma}
    \label{lem:r-drop}
    The probability of Step 5 in \Cref{alg:rounding} is at most $e^{-6}$.
\end{lemma}
\begin{proof}
    We use the following Chernoff bound: 
    If $X_1, \ldots, X_n$ are independent {0,1}-valued random variables with $X = \sum_{i=1}^n X_i$ such that $\mu = \E[X]$,
    then $\Pr[X > \mu + \beta] \leq \exp(-\frac{\beta^2}{2\mu + 2\beta / 3})$.
     
    Since the expected number of elements picked in $\pi_\ell$ is at most $8\cdot 2^\ell$, set $\beta = \mu = 8\cdot 2^\ell$, we have the the probability of more than $16 \cdot 2^\ell$ were picked is at most $\exp(\frac{-64\cdot 2^{2\ell}}{(64/3)2^\ell})\leq e^{-6}$.
\end{proof}

\begin{lemma}
    \label{lem:ori-approx}
    For any $q \leq 2\ceil{\log k}$, the expected cover time of any agent given by $\pi^q = \pi^q_1 \ldots \pi^q_{\ceil{\log n}}$ is at most $512 \cdot T^*$.
\end{lemma}
\begin{proof}
    For simplicity, fix $q$ and agent $a_i$, we may drop $q$ and $i$ from the notation. Let $\cE_{S_j, \ell}$ denote the event that $S_j$ is first covered in phase $\ell$. Consider an agent $a_i$. By linearity of expectation, we only focus on a set $S_j$. Then we have
    $$ \E[\cov(S_j, \pi^q)] \leq \sum_{\ell = \ceil{\log t^*_{S_j}}}^{\ceil{\log n}} 2 \cdot 16 \cdot 2^\ell \cdot \Pr[\cE_{S_j, \ell}]$$
    
    For each phase $\ell$, $S_j$ is not covered only if $K(S_j)$ elements from $S_j$ were not picked in $\pi^q_\ell$, or $\pi^q_\ell$ is empty.
    From \Cref{lem:r-Knap} and \Cref{lem:r-drop}, the probability that $S_j$ is not cover in any phase is at most $e^{-9/8} + e^{-6} < e^{-1}$. Therefore, we have 
    $$ \Pr[\cE_{S_j, \ell}] \leq \exp(-(\ell - \ceil{\log t^*_{S_j}}))$$
    
    Plugging in the inequality above, we have
    \begin{align*}
        \E[\cov(S_j, \pi^q)] &\leq \sum_{\ell = \ceil{\log t^*_{S_j}}}^{\ceil{\log n}} 2 \cdot 16 \cdot 2^\ell \cdot \exp(-(\ell - \ceil{\log t^*_{S_j}})) \\
        &\leq 256 \cdot t^*_{S_j}
    \end{align*}
    Then, by linearity of expectation, we get
    $$ \E[\sum_{j = 1}^m \cov(S_j, \pi^q)] = \sum_{j = 1}^m \E[\cov(S_j, \pi^q)] \leq 256 \cdot \sum_{j = 1}^m t^*_{S_j} \leq 512 \cdot T^*$$
\end{proof}

Now we are ready to prove the polylogarithmic approximation ratio.

\begin{lemma}
    The probability that the total cover time is at most $O(\log k) \cdot T^*$ for all agents is at least $1-\frac{1}{k}$, i.e., $\Pr[\max_{i \in [k]} \sum_{j=1}^m \cov(\sij, \pi) \leq 1024 \log k \cdot T^*] \geq 1 - \frac{1}{k}$, where $k$ is the number of agents.
\end{lemma}
\begin{proof}
    By union bound, it is sufficient to show that, for every agent $a_i$, the probability that the total cover time of agent $a_i$ is at least $1 - \frac{1}{k^2}$.
    %$\Pr[\sum_{j=1}^m \cov(\sij, \pi) \leq 1024 \log k \cdot T^*] \geq 1 - \frac{1}{k^2}$. 
    From \Cref{lem:ori-approx}, for every $q$, the expected total cover time given by $\pi^q$ is at most $512 \cdot T^*$.
    %$$ \E[\sum_{j=1}^m \cov(\sij, \pi^q_1 ... \pi^q_{\ceil{\log n}})] \leq 512 \cdot T^* $$.
    Then, by Markov's inequality, we have 
    $$ \Pr[\sum_{j=1}^m \cov(\sij, \pi^q) \leq 1024 \cdot T^*] \geq 1/2$$
    
    Recall that $\pi = \pi_1^1 \ldots \pi_1^{\log k} \ldots \pi_{\lceil \log n \rceil}^1 \ldots \pi_{\lceil \log n \rceil}^{\log k}$, then, for any agent $a_i$, $\sum_{j=1}^m \cov(\sij, \pi) > 1024 \log k \cdot T^*$ happens only if $\sum_{j=1}^m \cov(\sij, \pi^q) > 1024 \cdot T^*$ for all $\ceil{2\log k}$ independent rounding. Therefore, we get 
    $$ \Pr[\sum_{j=1}^m \cov(\sij, \pi) \leq 1024 \log k \cdot T^*] \geq 1 - 2^{-2\ceil{\log k}} \geq 1 - 1/k^2$$ 
    
%    Let $\cE_{\sij, \ell}$ denote the event that $\sij$ is first covered in phase $\ell$.
    
%    $$ \Pr[\cE_{\sij, \ell}] \leq \exp(-\log k (\ell - \log (10 \log k \cdot t^*_{\sij}))$$
%    \begin{align*}
%        \Pr[\cov(\sij, \pi) \leq 10 \log k \cdot t^*_{\sij}] &\geq 1 - \sum_{\ell = \lceil \log t^*_{\sij} + \log \log k + O(1) \rceil} ^{\lceil \log n \rceil + 1} %\Pr[\cE_{\sij, \ell}] \\
%        &\geq 1 - 1 / k^3
%    \end{align*}
    %
\end{proof}
\section{Experiments}

This section investigates the empirical performance of our algorithms. We seek to show that the theory is predictive of practice on real data. We give experimental results for the min-max optimal decision tree over multiple agents. 
At the high level, there is a set of objects where each object is associated with an attribute vector. We can view the data as a table with the objects as rows and attributes as columns. Each agent $i$ has a target object $o_i$ sampled from a different distribution over the objects. Our goal is to order the columns to find all agents' targets as soon as possible. When column $c$ is chosen, for each agent $i$,  rows that have a different value from $o_i$ in column $c$ are discarded. The target object $o_i$ is identified when all the other objects are inconsistent with $o_i$ in one of the columns probed.

%The agents aim to find objects with certain attributes in the table, and our goal is to order the columns such that the agents can find their targets as soon as possible when they filter the columns sequentially in the order. We first introduce the data sets and how we construct instances.

\paragraph{Data Preparation.} 

In the experiments, three public data sets are considered: MFCC data set\footnote{\url{https://archive.ics.uci.edu/ml/datasets/Anuran+Calls+\%28MFCCs\%29}}, PPPTS data set\footnote{\url{https://archive.ics.uci.edu/ml/datasets/Physicochemical+Properties+of+Protein+Tertiary+Structure\#}}, and CTG data set\footnote{\url{https://archive.ics.uci.edu/ml/datasets/Cardiotocography}}. These are the data sets in the field of life science. Each one is in table format and the entries are real-valued.
 The sizes of the three data sets are $7195 \times 22$, $45730 \times 9$, and $2126 \times 23$ respectively, where $h \times n$ implies $h$ rows and $n$ columns. 
%Due to space, we defer their detailed descriptions to \cref{app:data}. 

The instances are constructed as follows. 
We first discretize real-valued entries, so entries in each column can have at most ten different values. This is because the objects' vectors could have small variations, and we seek to make them more standardized.
%To avoid the case that most rows can be identified easily by any attribute, we perform the following rounding. 
%For each column, say $a$ and $b$ are the minimum value and the maximum value that occurred in the column, respectively. We partition the interval $[a,b]$ into ten equal-length sub-intervals and round each entry in this column to the upper bound of the sub-interval that it belongs. 
%After rounding, different rows may share the same attribute vector.
Let the ground set of elements $U$ be all the columns for a table, and view each row $j$ as an object. Create a submodular function $f_j$ for each object $j$: For a subset $S\subseteq U$, $f_j(S)$ represents the number of rows that  have different attributes from row $j$ after checking the columns in $S$. If $f_j(S)=f(U)$, the object in row $j$ can be completely distinguished from other objects by checking columns in $S$. We then normalize functions to have a range of $[0, 1]$.
%This function is covered, that is, the agent succeeds in finding the rows with certain attributes.
Note that the functions are created essentially in the same way they are in the reduction of the optimal decision tree to submodular ranking, as discussed in Section~\ref{sec:intro}. All agents have the same table, that is, the same objects, functions, and columns. We construct a different weight vector for each agent, where each entry has a value sampled uniformly at random from $[1,100]$. %Note that this is equivalent to an object being sampled with probability in proportion to a weight randomly chosen from $[1,100]$.
In the following, we use $K$ and $M$ to denote the number of agents and the number of functions, respectively.

%We set up $K$ agents and sample $M$ functions for each agent. For each agent-function pair, sample a number uniformly from $[1,100]$ to serve as the weight. 
%\sungjin{What is $M$?}

%\sungjin{2(a). data to Dataset. 2(b) K to $K$, Number of Agents. 2(c) M  to $M$ Number of Functions Per Agent. 2(d) $K = M $. Also I think it's better to use full names. Put them only in b c or d. e.g. in (c), R: Random, G: Greedy, NG: Normalized Greedy. BAG: balanced adaptive Greedy}

\begin{figure*}[t]
\centering
\subfigure[]{
%\begin{minipage}{.32\textwidth}
    \centering
    \includegraphics[width=0.45\textwidth]{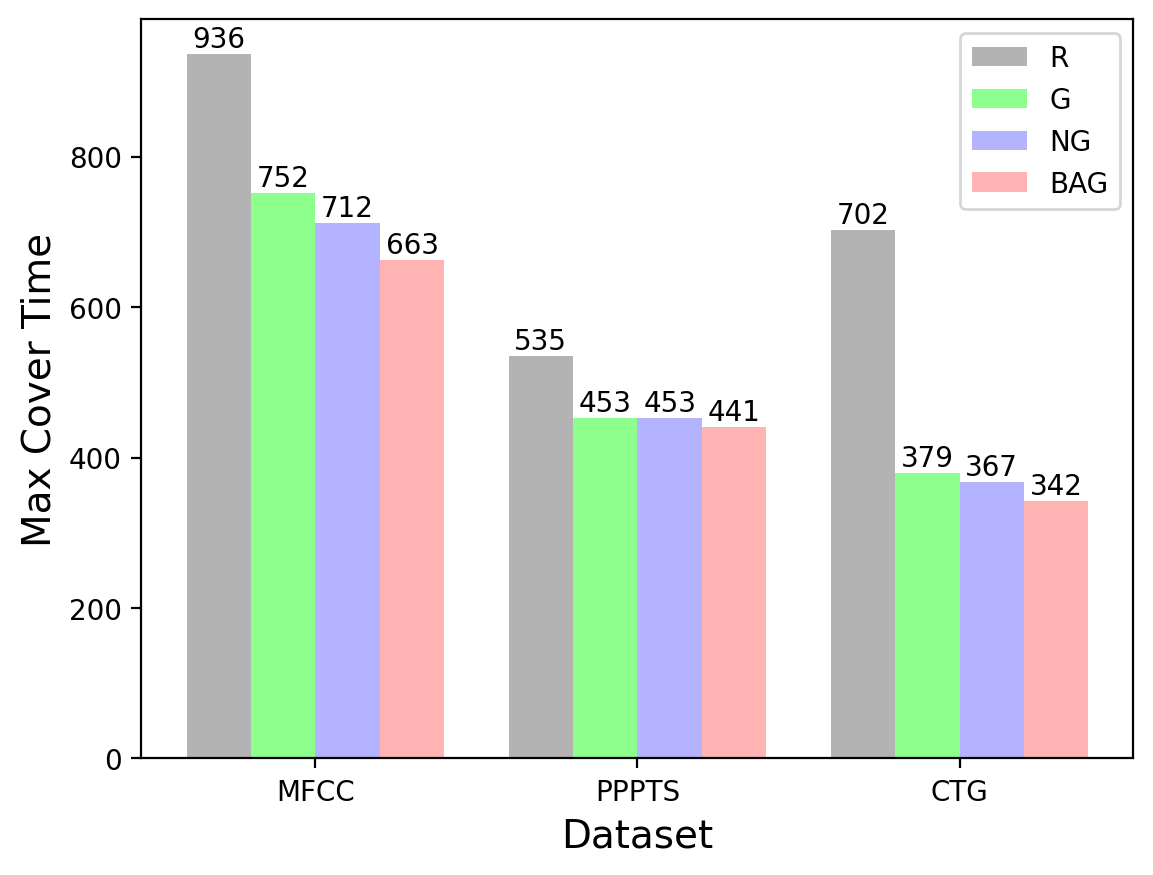}
    \label{fig:all_datasets}
   % \end{minipage} 
    }
\subfigure[]{
%\begin{minipage}{.32\textwidth}
    \centering
    \includegraphics[width=0.45\textwidth]{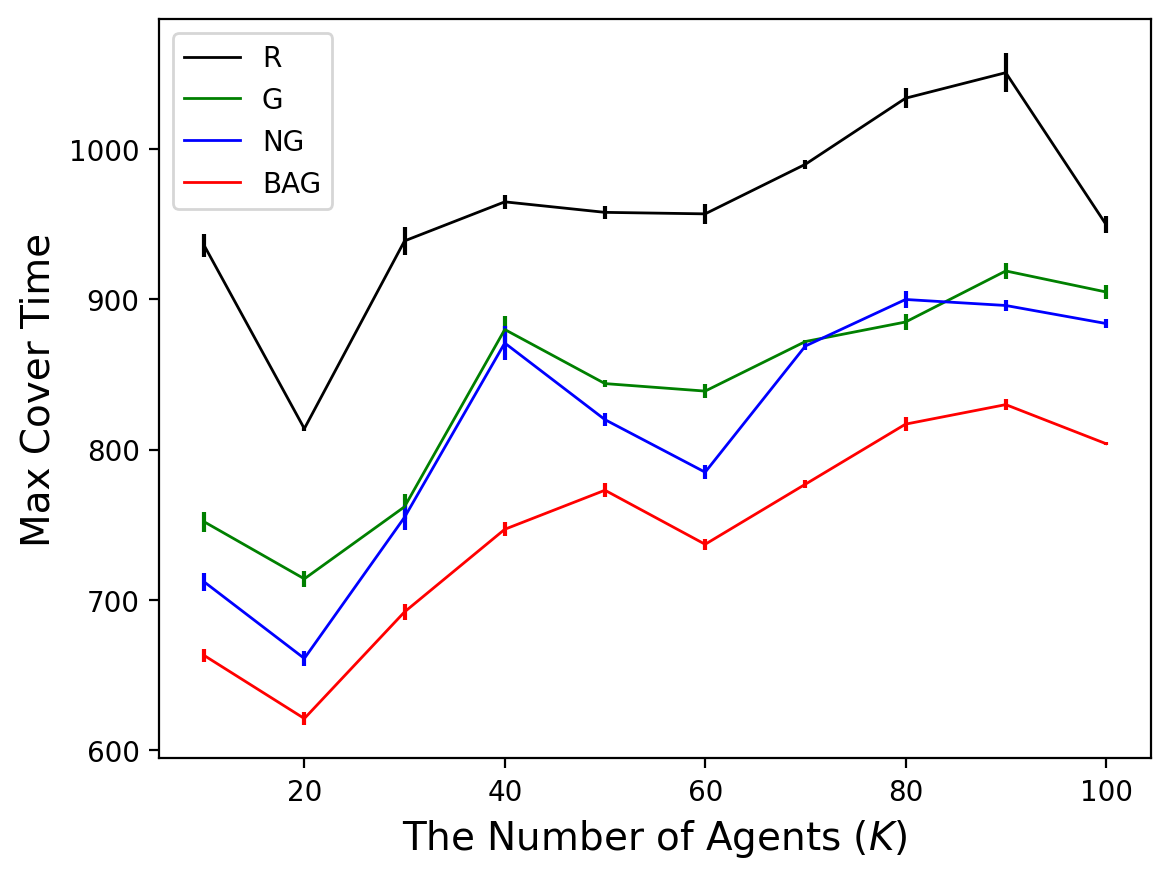}
    \label{fig:MFCC_K}
   % \end{minipage} 
    }
    
\subfigure[]{
%\begin{minipage}{.32\textwidth}
    \centering
    \includegraphics[width=0.45\textwidth]{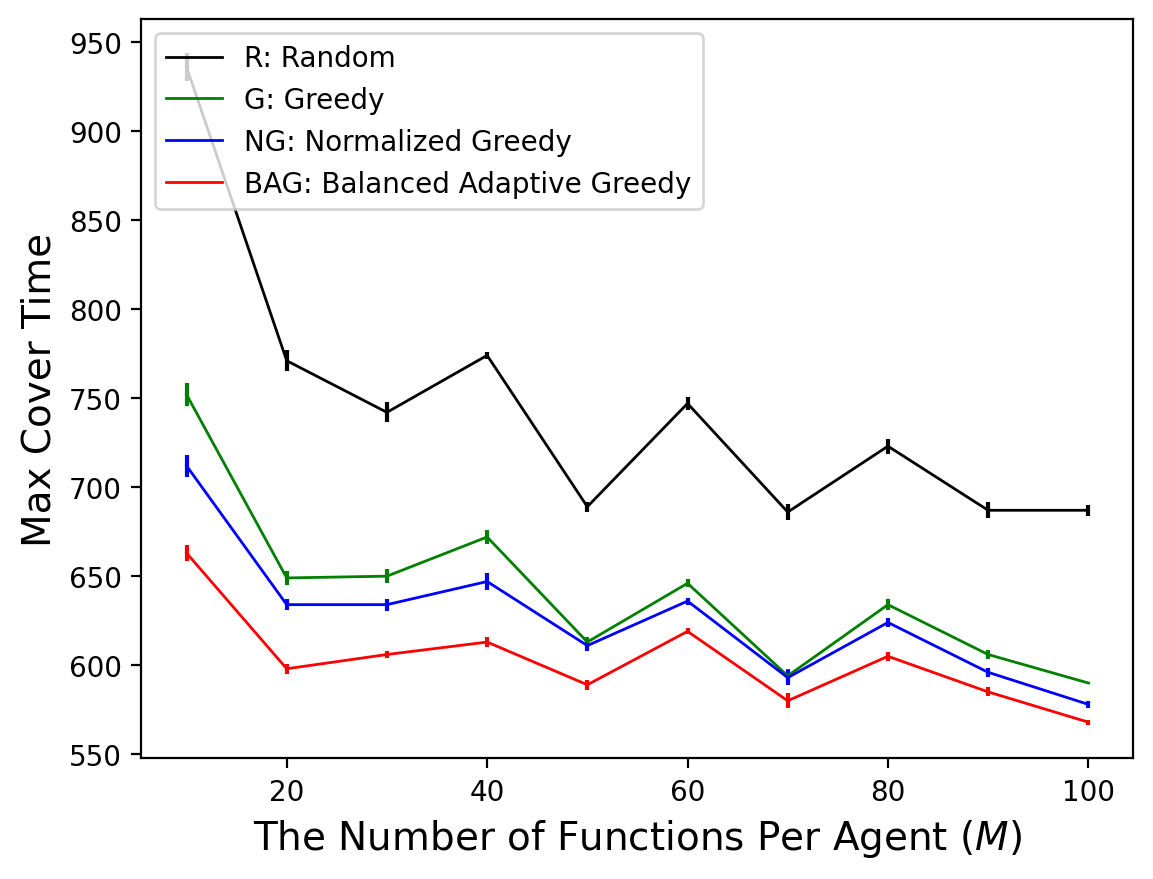}
  %  \caption{Runtime on Type Model instances with exponentially increasing noise}
    \label{fig:MFCC_M}
    %\end{minipage}
    }
    %\caption{}
\subfigure[]{  
    %\begin{minipage}{.32\textwidth}
    \centering
    \includegraphics[width=0.45\textwidth]{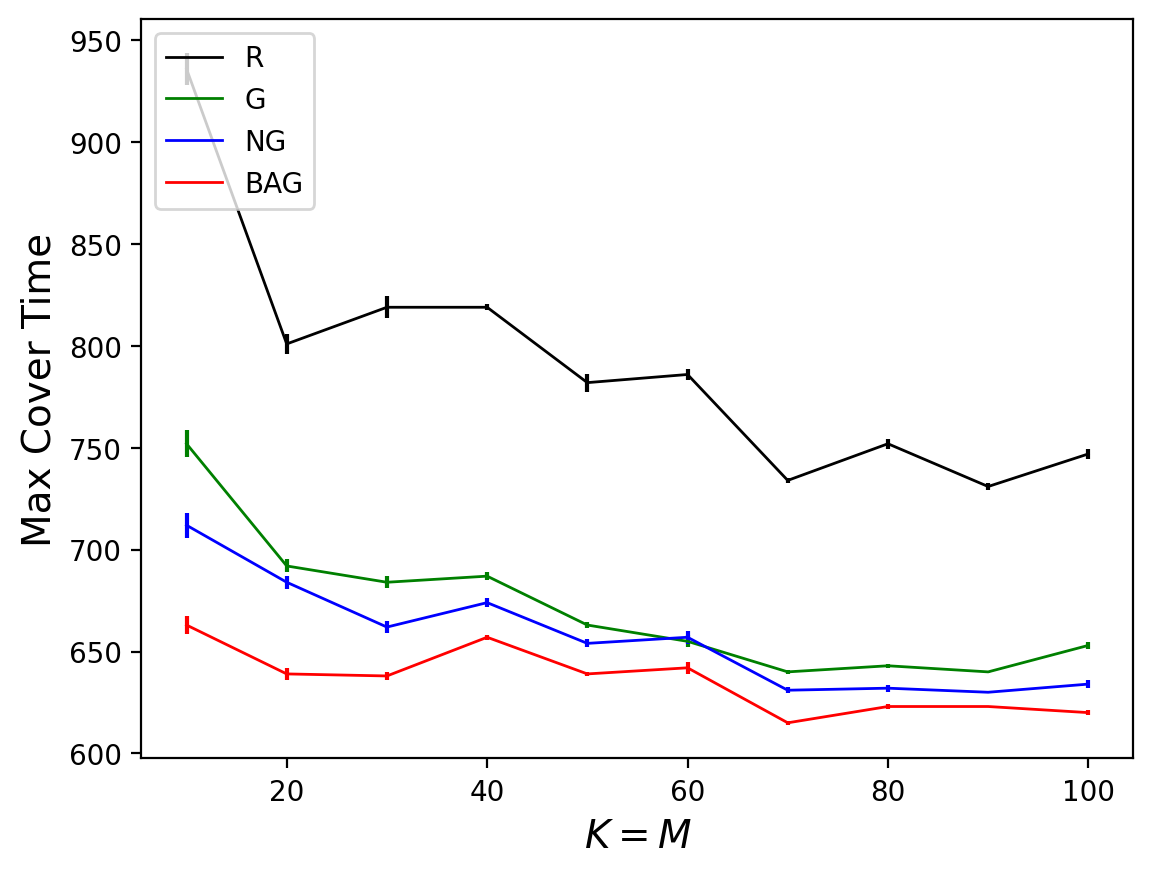}
 %   \caption{Runtime results for online setting on the type model}
    \label{fig:MFCC_KM}
    %\end{minipage}
    }
    \caption{\cref{fig:all_datasets} shows the results on different datasets when both the number of agents $K$ and the number of functions per agent $M$ are 10. Others show the performance of algorithms on the MFCC dataset when $K$ and $M$ vary. In \cref{fig:MFCC_K}, we fix $M=10$ and increase $K$, while in \cref{fig:MFCC_M}, we fix $K=10$ and increase $M$. Finally, \cref{fig:MFCC_KM} shows the algorithms' performance when $K$ and $M$ are set to be the same value and increase together.
    }
    \label{fig:MFCC}
\end{figure*}

\paragraph{Baseline Algorithms and Parameter Setting.} We refer to our main  \cref{alg:comb} as balanced adaptive greedy (BAG). We also refer to the naive adaptation of the algorithm proposed by \cite{DBLP:conf/soda/AzarG11} as normalized greedy (NG). For full description of NG, see \cref{sec:ng}.
%
%We refer to the two algorithms proposed in \cref{sec:ng} and \cref{sec:BAG} as the normalized greedy (NG) algorithm and the balanced adaptive greedy (BAG) algorithm. 
The algorithms are compared to two natural baseline algorithms. One is called the random (R) algorithm, which directly outputs a random permutation of elements. The other is the greedy (G) algorithm, which selects the element that maximizes the total increment of all $K\cdot M$ functions each time. Notice that  \cref{alg:comb}, the decreasing ratio of the sequence in algorithm BAG is set to be $2/3$, but in practice, this decreasing ratio is flexible. In the experiments, we test the performance of algorithm BAG with decreasing ratios in $[0,0.05,0.1,\ldots,0.95,1]$ and pick the best decreasing ratio. 

We conduct the experiments\footnote{The code is available at \url{https://github.com/Chenyang-1995/Min-Max-Submodular-Ranking-for-Multiple-Agents}} on a machine running Ubuntu 18.04 with an i7-7800X CPU and 48 GB memory. We investigate the performance of algorithms on different data sets under different values of $K$ and $M$. The results are averaged over four runs. The data sets give the same trend. Thus, we first show the algorithms' performance with $K=M=10$ on the three data sets and only present the results on the MFCC data set when $K$ and $M$ vary in~\cref{fig:MFCC}. 
%We now show the performance of algorithms on the PPPTS data set (\cref{fig:PPPTS}) and the CTG data set (\cref{fig:CTG}). The figures show the same trends. In addition, we also investigate the empirical results of these algorithms if the objective is the average cover time of agents, rather than the maximum cover time. We see that even in this case, our balanced adaptive greedy method still outperforms other algorithms on different data sets (\cref{fig:MFCC(avg)}, \cref{fig:PPPTS(avg)} and \cref{fig:CTG(avg)}).
The results on the other two data sets appear in~\cref{app:other_data}.

\paragraph{Empirical Discussion.} From the figures, we see that the proposed balanced adaptive greedy algorithm always obtains the best performance for all datasets and all values of $K$ and $M$. Moreover, \cref{fig:MFCC_K} shows that as $K$ increases, the objective value of each algorithm generally increases, implying that the instance becomes harder. In these harder instances, algorithm BAG has a more significant improvement over other methods.
Conversely, \cref{fig:MFCC_M} indicates that we get easier instances as $M$ increases because all the curves generally give downward trends. In this case, although the benefit of our balancing strategy becomes smaller, algorithm BAG still obtains the best performance.

%each column (attribute) is an element. The experiments use the following way to generate functions for agents.  
\section{Conclusion}

The paper is the first to study the submodular ranking problem in the presence of multiple agents. The objective is to minimize the maximum cover time of all agents, i.e., optimizing the worst-case fairness over the agents. This problem generalizes to designing optimal decision trees over multiple agents and also captures other practical applications. 
By observing the shortfall of the existing techniques, we introduce a new algorithm, balanced adaptive greedy. Theoretically, the algorithm is shown to have strong approximation guarantees. The paper shows empirically that the theory is predictive of experimental performances. Balanced adaptive greedy is shown to outperform strong baselines in the experiments, including the most natural greedy strategies.

The paper gives a tight approximation algorithm for generalized min-sum set cover on multiple agents, which is a special case of our model. 
The upper bound shown in this paper matches the lower bound introduced. 
The tight approximation for the general case is left as an interesting open problem.
Beyond the generalized min-sum set cover problem, another special case of our problem is also interesting in which the monotone submodular functions of each agent are the same.
Observing the special case above also captures the problem of Optimal Decision Tree with Multiple Probability Distribution.
Thus, improving the approximation for this particular case will be interesting.

\section*{Acknowledgments}
Chenyang Xu was supported in part by Science and Technology Innovation 2030 –``The Next Generation of Artificial Intelligence" Major Project No.2018AAA0100900.
Qingyun Chen and Sungjin Im were supported in part by 
NSF CCF-1844939 and CCF-2121745. Benjamin Moseley was supported in part by  a Google Research Award, an Infor Research Award, a Carnegie Bosch Junior Faculty Chair and NSF grants CCF-1824303,  CCF-1845146, CCF-1733873 and CMMI-1938909.
We thank the anonymous reviewers for their insightful comments and suggestions.

\clearpage

\printbibliography

\newpage
\appendix
\section{More Experimental Results}
\label{app:more_exp}

\subsection{Data Set Descriptions}
\label{app:data}

The experments are implemented on three public data sets: mel-frequency cepstral coefficients of anuran calls (MFCC) data set\footnote{\url{https://archive.ics.uci.edu/ml/datasets/Anuran+Calls+\%28MFCCs\%29}}, physicochemical properties of protein tertiary structures (PPPTS) data set\footnote{\url{https://archive.ics.uci.edu/ml/datasets/Physicochemical+Properties+of+Protein+Tertiary+Structure\#}}, and diagnostic features of cardiotocograms (CTG) data set\footnote{\url{https://archive.ics.uci.edu/ml/datasets/Cardiotocography}}. 

The first MFCC data set was created by 60 audio records belonging to 4 different families, 8 genus, and 10 species. It has 7195 rows and 22 columns, where each row corresponds to one specimen (an individual frog) and each column is an attribute. 
The second PPPTS data set is a data set of Physicochemical Properties of Protein Tertiary Structure. It is taken from Critical Assessment of protein Structure Prediction (CASP), which is a worldwide experiment for protein structure prediction taking place every two years since 1994. 
The data set consists of 45730 rows and 9 columns. The third CTG data set consists of measurements of fetal heart rate (FHR) and uterine contraction (UC) features on cardiotocograms classified by expert obstetricians. It has 2126 rows and 23 columns, corresponding to 2126 fetal cardiotocograms and 23 diagnostic features.

\subsection{Results on Other Data Sets}
\label{app:other_data}

\begin{figure*}[htbp]
\centering
\subfigure[]{
%\begin{minipage}{.32\textwidth}
    \centering
    \includegraphics[width=0.3\textwidth]{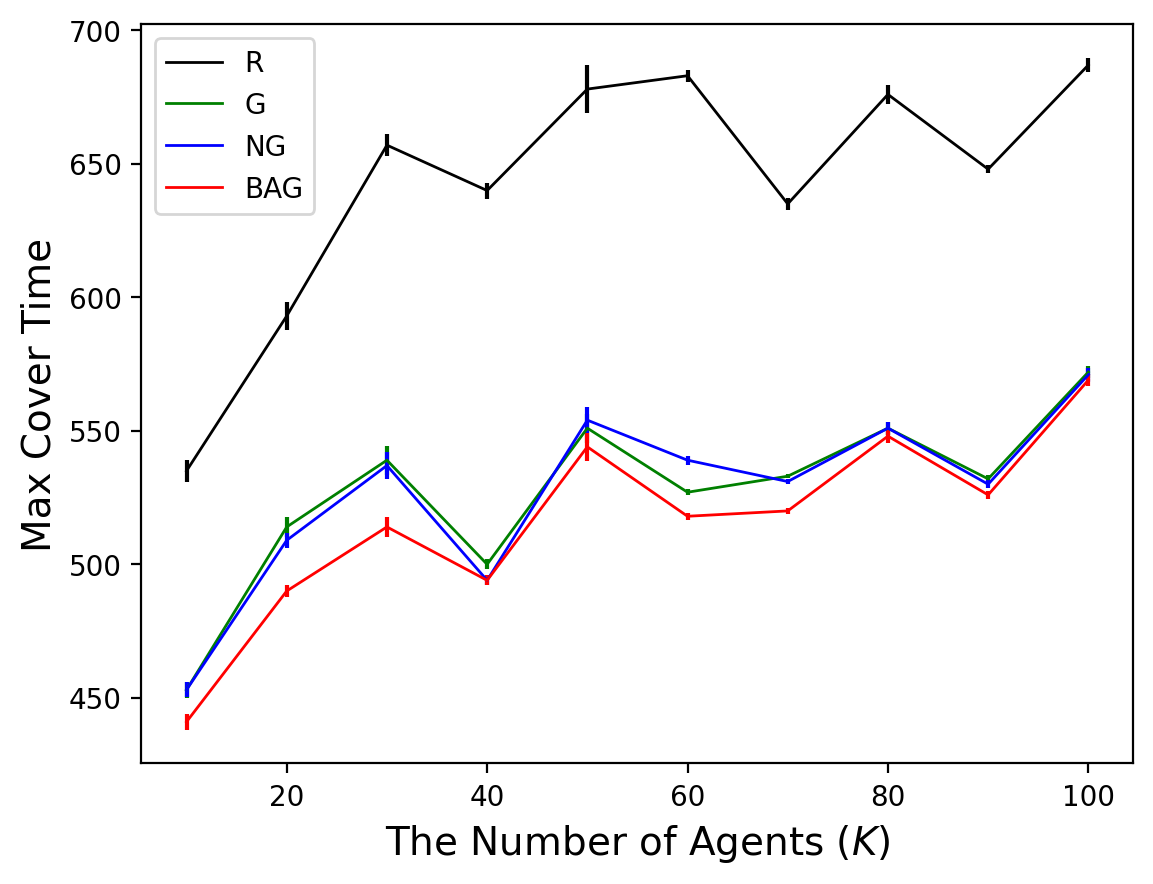}
    \label{fig:PPPTS_K}
   % \end{minipage} 
    }
\subfigure[]{
%\begin{minipage}{.32\textwidth}
    \centering
    \includegraphics[width=0.3\textwidth]{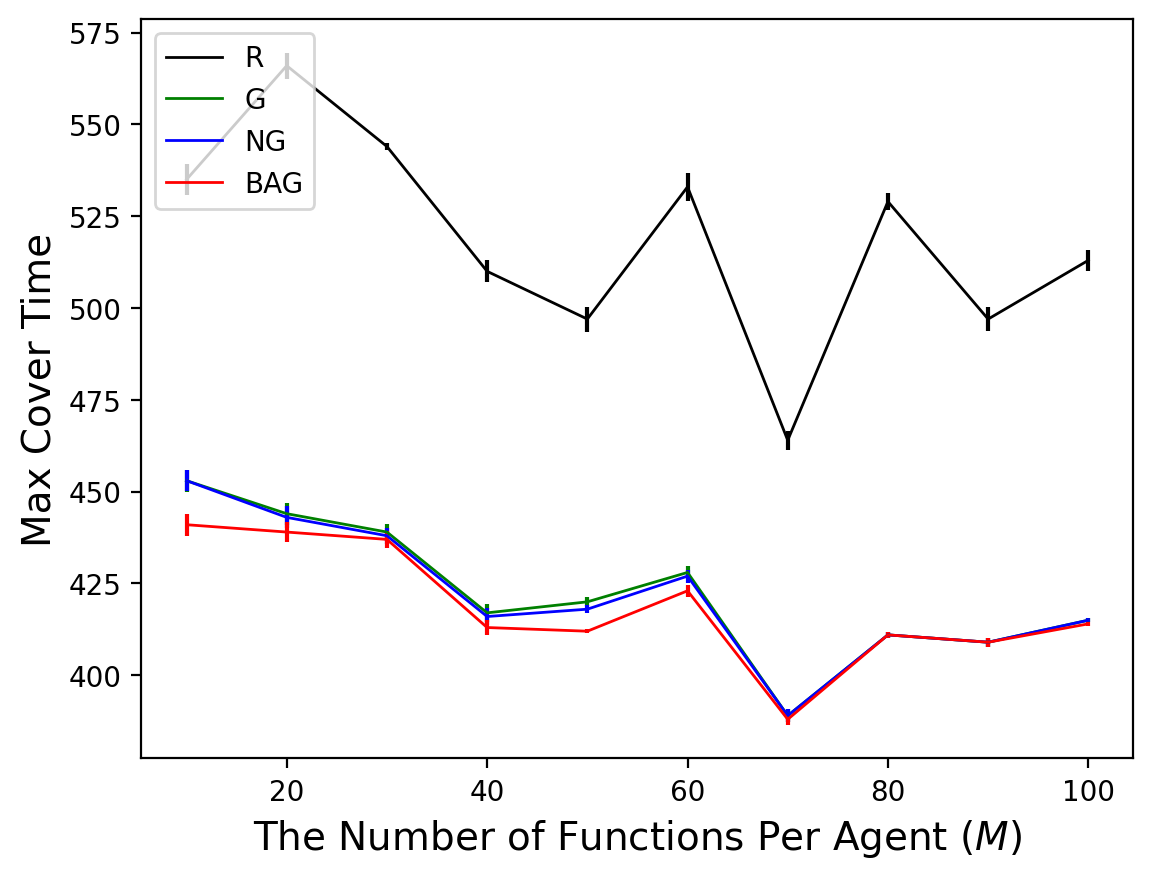}
  %  \caption{Runtime on Type Model instances with exponentially increasing noise}
    \label{fig:PPPTS_M}
    %\end{minipage}
    }
    %\caption{}
\subfigure[]{  
    %\begin{minipage}{.32\textwidth}
    \centering
    \includegraphics[width=0.3\textwidth]{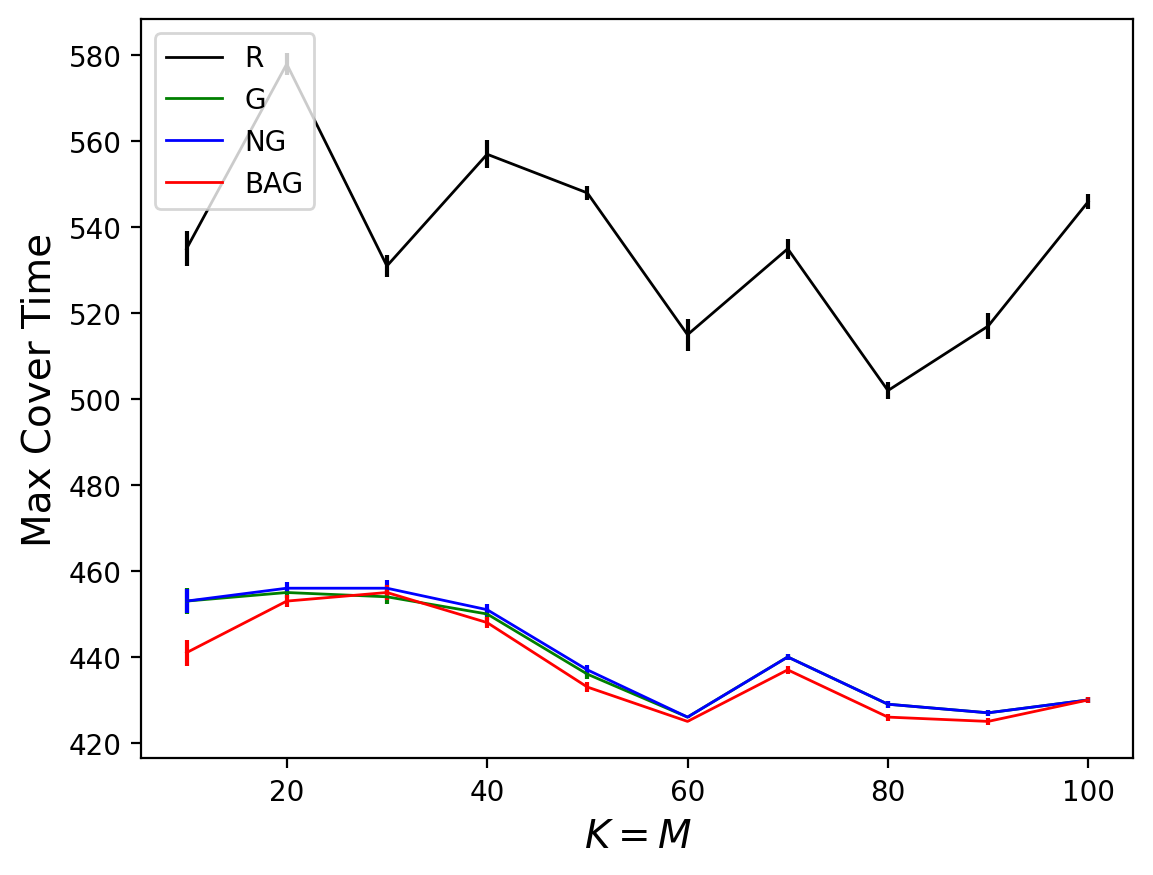}
 %   \caption{Runtime results for online setting on the type model}
    \label{fig:PPPTS_KM}
    %\end{minipage}
    }
    \caption{The performance of algorithms on the PPPTS dataset when $K$ and $M$ vary. 
    }
    \label{fig:PPPTS}
\end{figure*}

\begin{figure*}[htbp]
\centering
\subfigure[]{
%\begin{minipage}{.32\textwidth}
    \centering
    \includegraphics[width=0.3\textwidth]{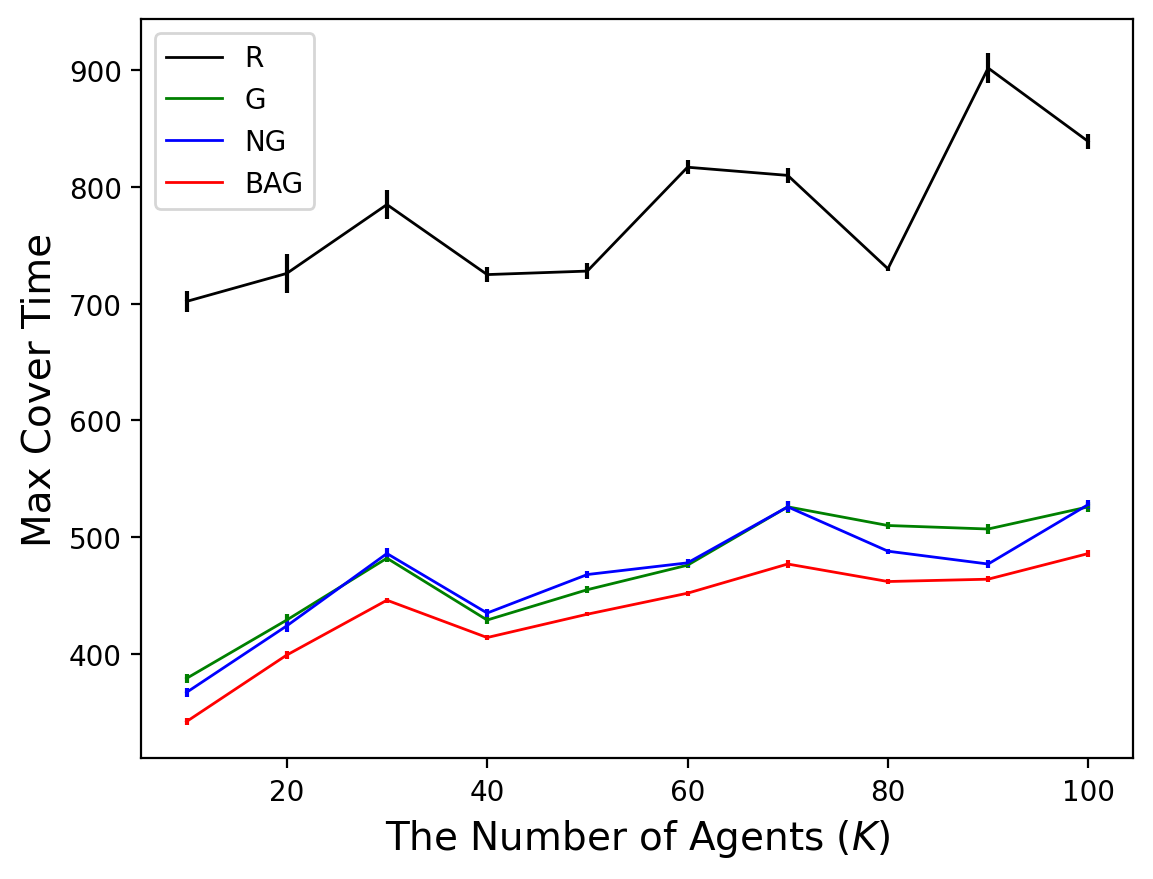}
    \label{fig:CTG_K}
   % \end{minipage} 
    }
\subfigure[]{
%\begin{minipage}{.32\textwidth}
    \centering
    \includegraphics[width=0.3\textwidth]{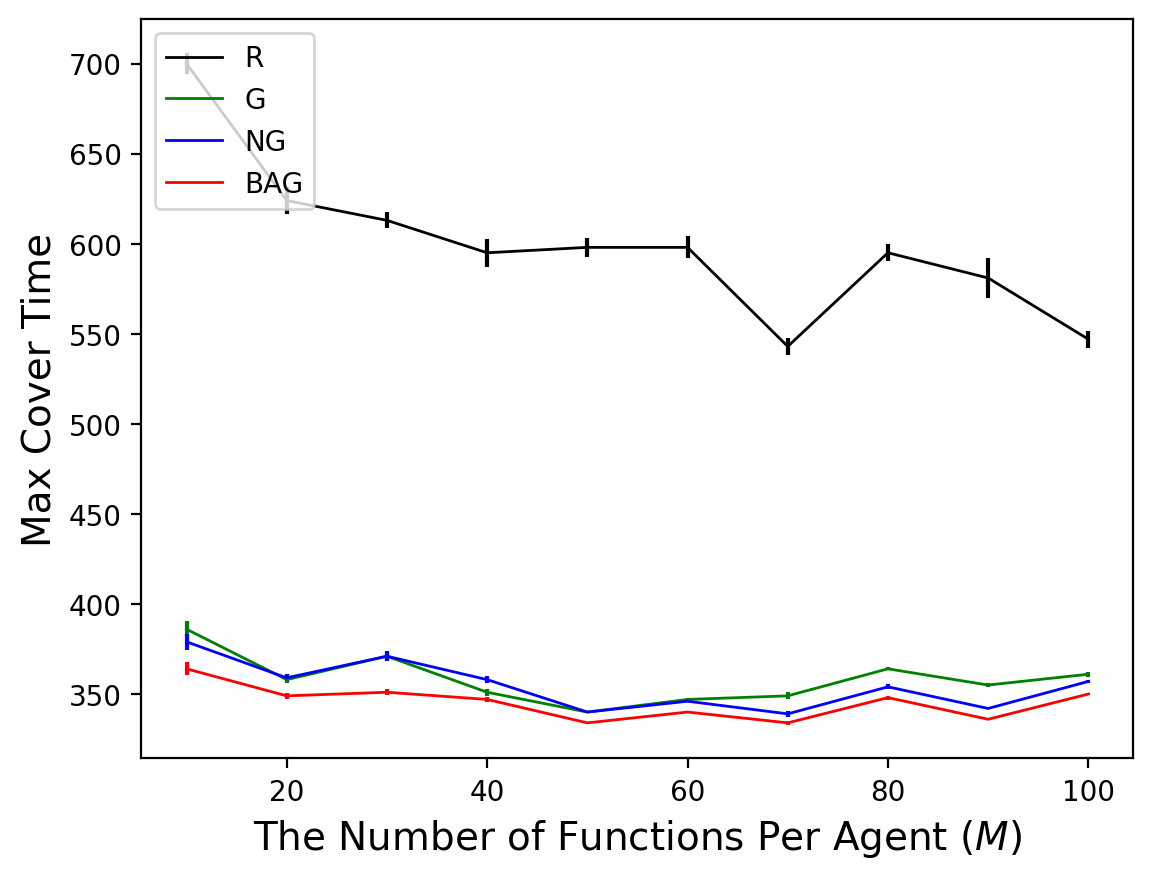}
  %  \caption{Runtime on Type Model instances with exponentially increasing noise}
    \label{fig:CTG_M}
    %\end{minipage}
    }
    %\caption{}
\subfigure[]{  
    %\begin{minipage}{.32\textwidth}
    \centering
    \includegraphics[width=0.3\textwidth]{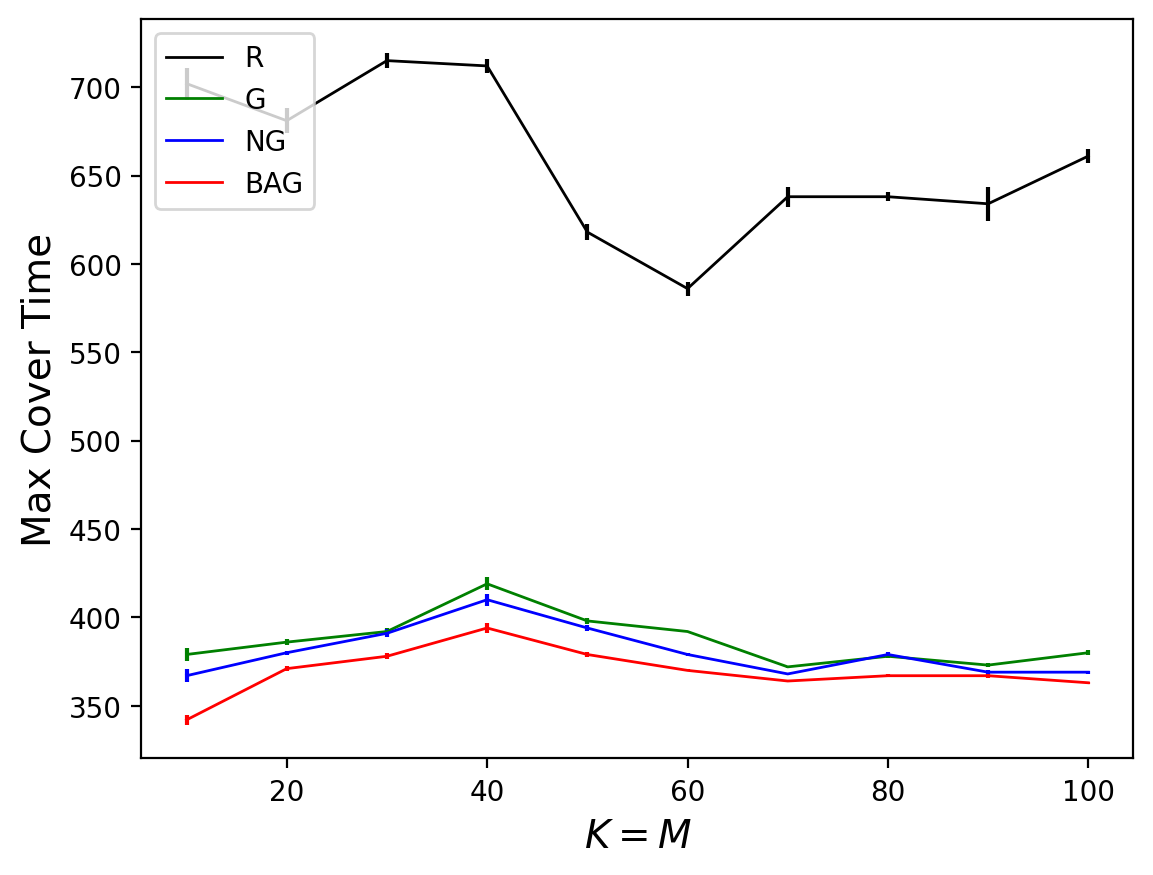}
 %   \caption{Runtime results for online setting on the type model}
    \label{fig:CTG_KM}
    %\end{minipage}
    }
    \caption{The performance of algorithms on the CTG dataset when $K$ and $M$ vary. 
    }
    \label{fig:CTG}
\end{figure*}

\begin{figure*}[htbp]
\centering
\subfigure[]{
%\begin{minipage}{.32\textwidth}
    \centering
    \includegraphics[width=0.3\textwidth]{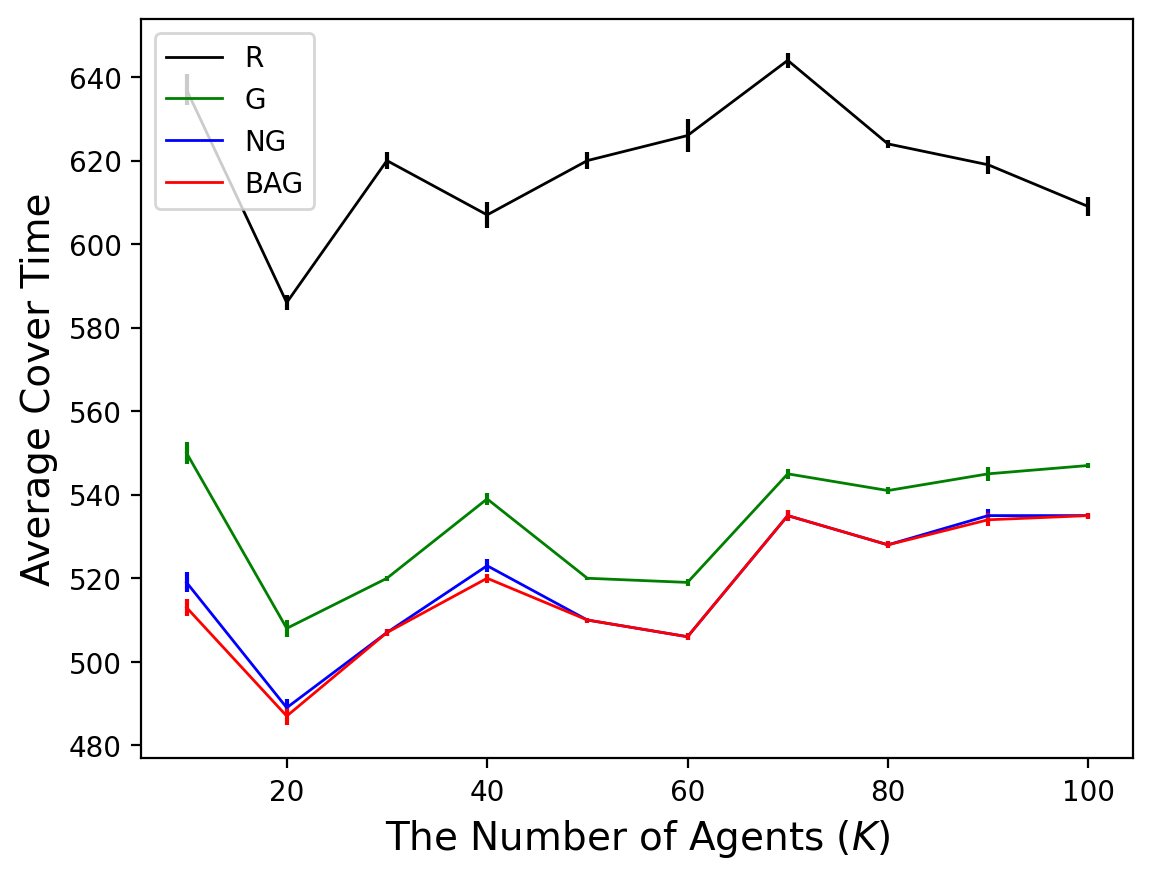}
    \label{fig:MFCC(avg)_K}
   % \end{minipage} 
    }
\subfigure[]{
%\begin{minipage}{.32\textwidth}
    \centering
    \includegraphics[width=0.3\textwidth]{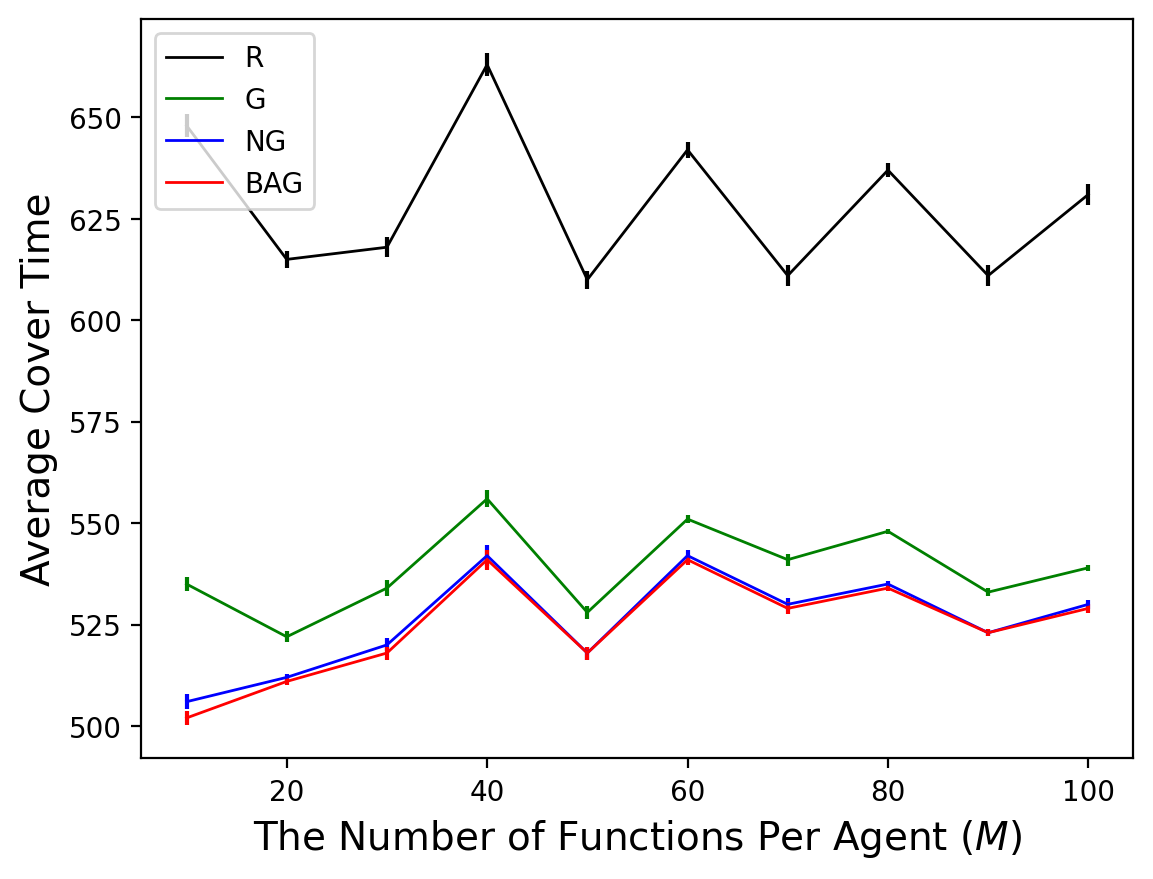}
  %  \caption{Runtime on Type Model instances with exponentially increasing noise}
    \label{fig:MFCC(avg)_M}
    %\end{minipage}
    }
    %\caption{}
\subfigure[]{  
    %\begin{minipage}{.32\textwidth}
    \centering
    \includegraphics[width=0.3\textwidth]{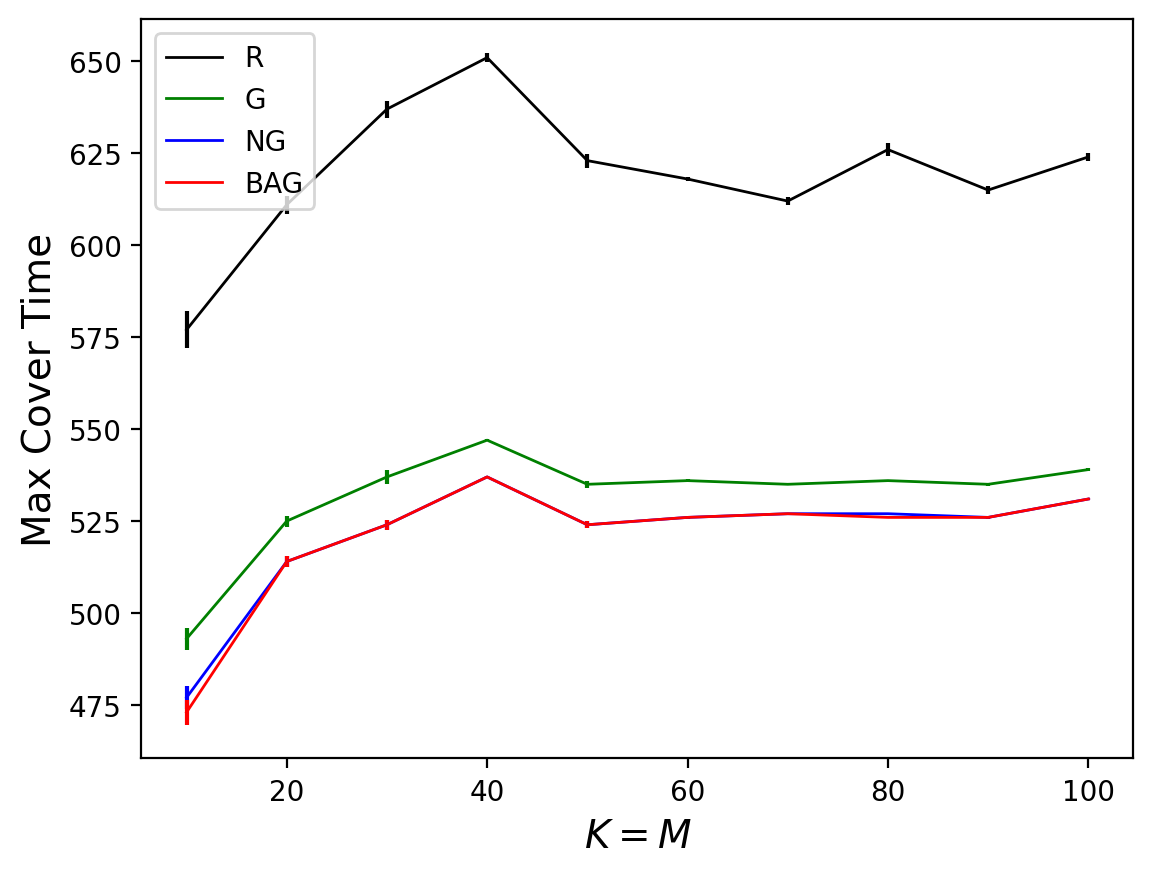}
 %   \caption{Runtime results for online setting on the type model}
    \label{fig:MFCC(avg)_KM}
    %\end{minipage}
    }
    \caption{The algorithms' average cover times of agents on the MFCC dataset when $K$ and $M$ vary. 
    }
    \label{fig:MFCC(avg)}
\end{figure*}

\begin{figure*}[htbp]
\centering
\subfigure[]{
%\begin{minipage}{.32\textwidth}
    \centering
    \includegraphics[width=0.3\textwidth]{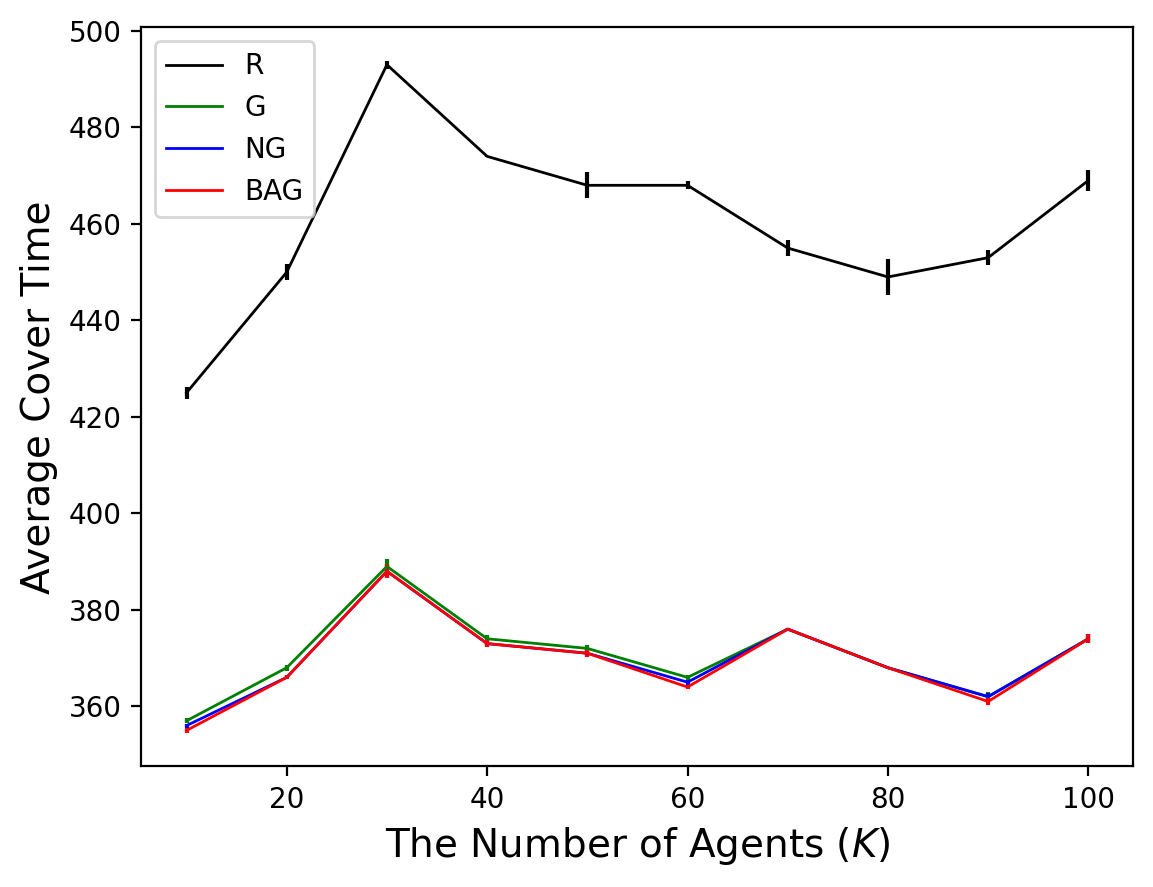}
    \label{fig:PPPTS(avg)_K}
   % \end{minipage} 
    }
\subfigure[]{
%\begin{minipage}{.32\textwidth}
    \centering
    \includegraphics[width=0.3\textwidth]{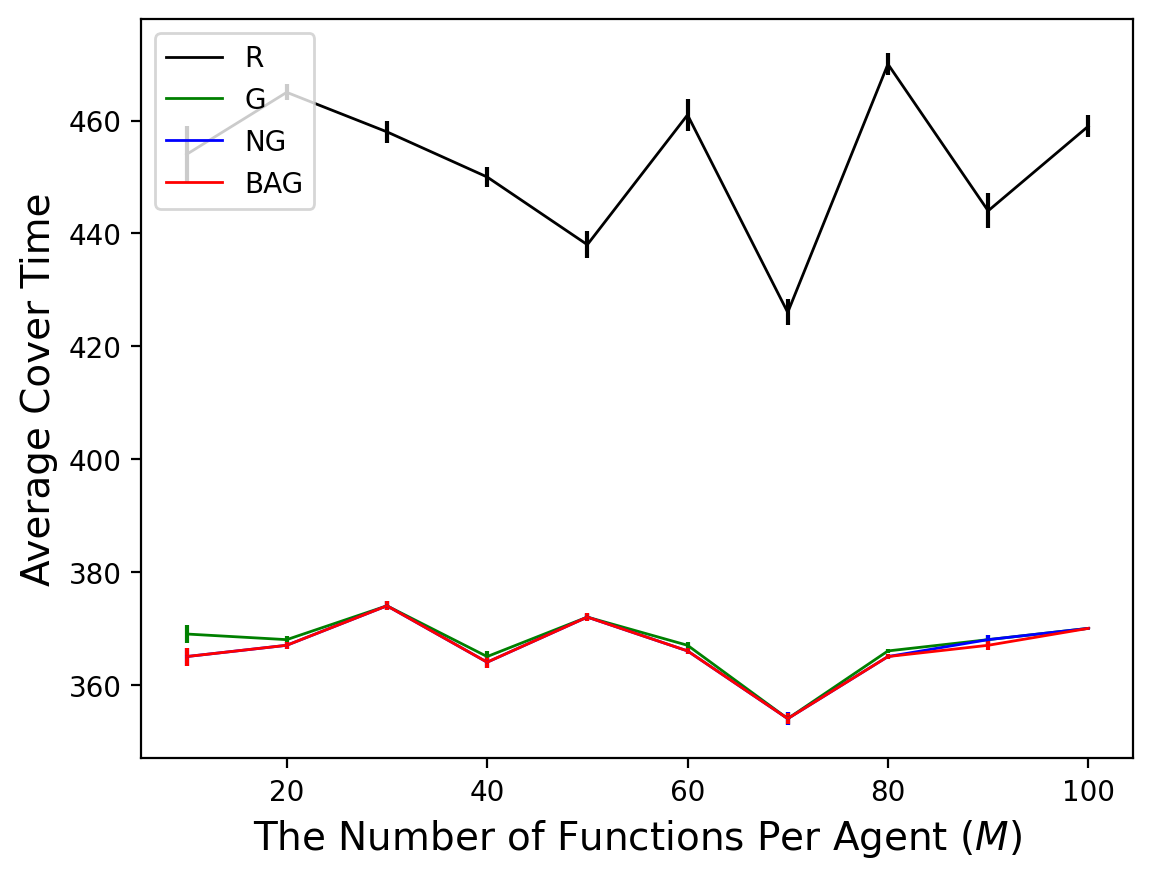}
  %  \caption{Runtime on Type Model instances with exponentially increasing noise}
    \label{fig:PPPTS(avg)_M}
    %\end{minipage}
    }
    %\caption{}
\subfigure[]{  
    %\begin{minipage}{.32\textwidth}
    \centering
    \includegraphics[width=0.3\textwidth]{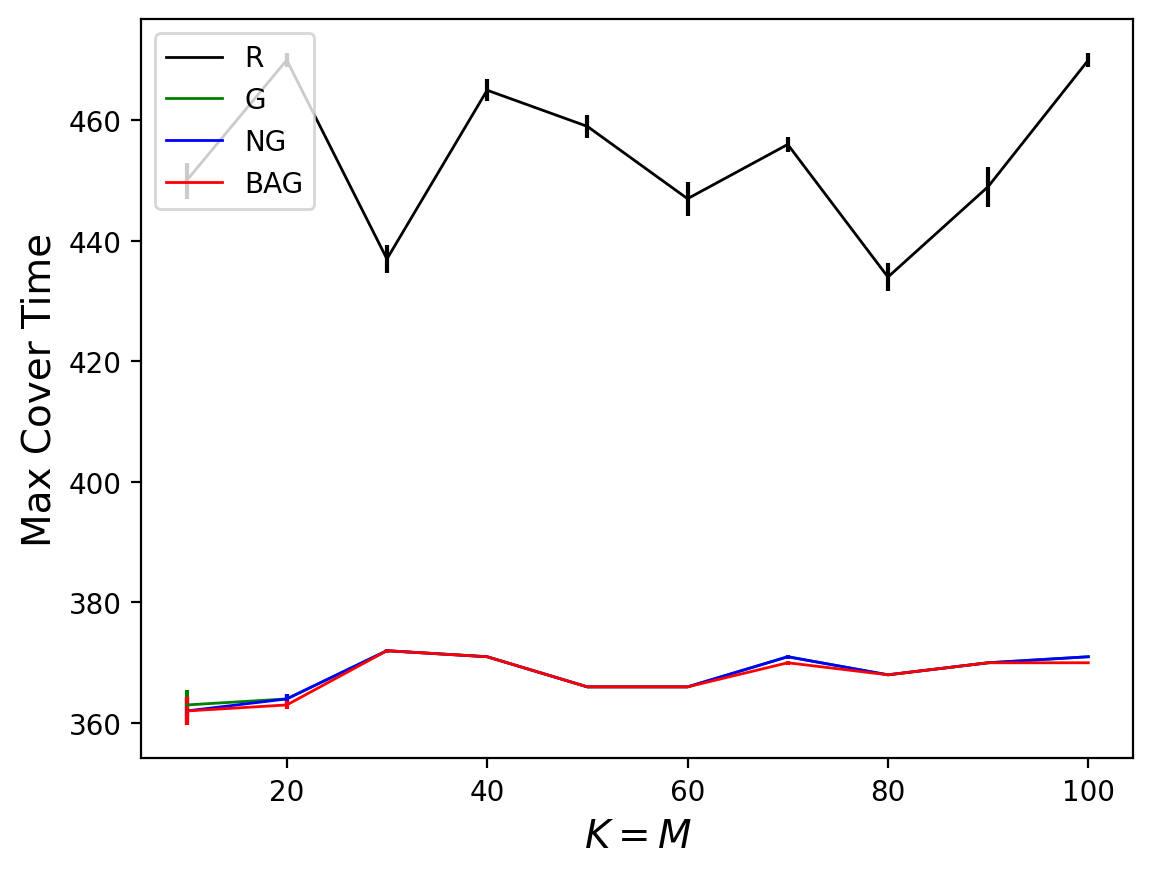}
 %   \caption{Runtime results for online setting on the type model}
    \label{fig:PPPTS(avg)_KM}
    %\end{minipage}
    }
    \caption{The algorithms' average cover times of agents on the PPPTS dataset when $K$ and $M$ vary. 
    }
    \label{fig:PPPTS(avg)}
\end{figure*}

\begin{figure*}[htbp]
\centering
\subfigure[]{
%\begin{minipage}{.32\textwidth}
    \centering
    \includegraphics[width=0.3\textwidth]{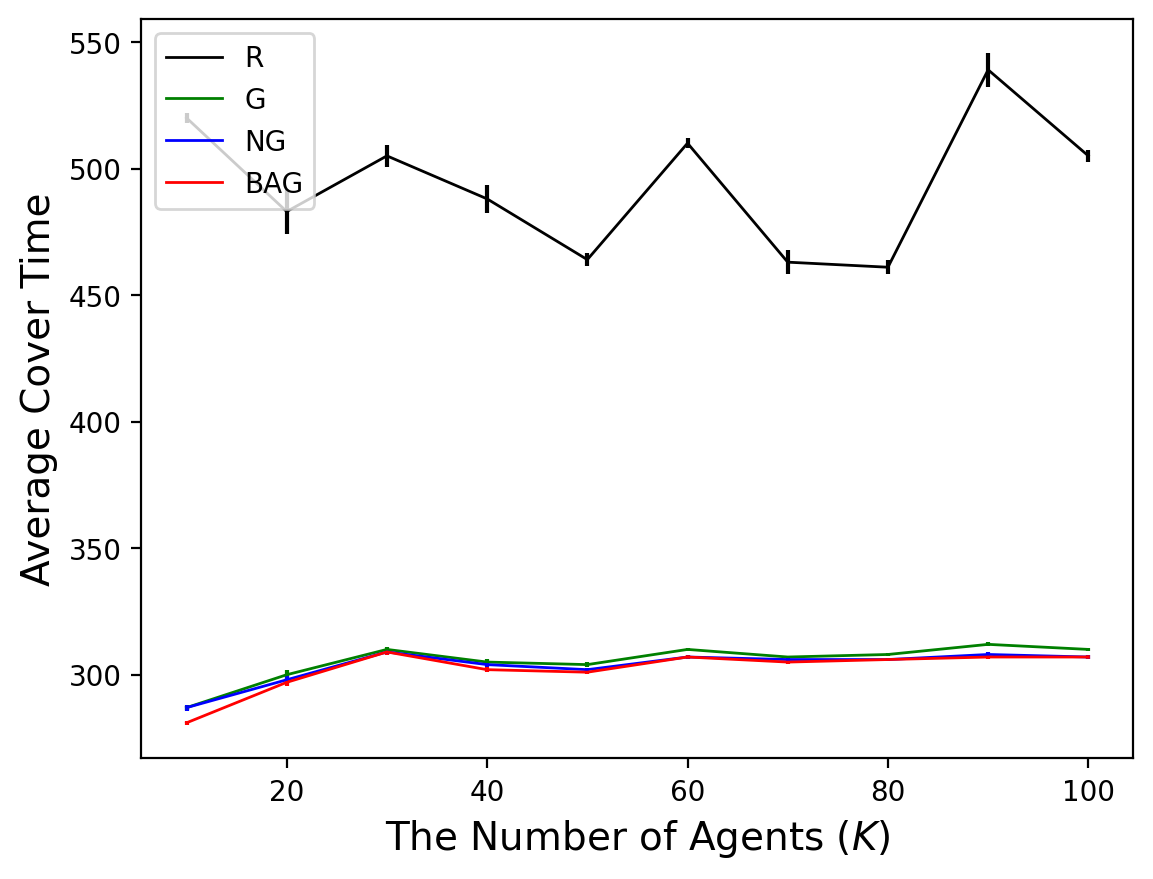}
    \label{fig:CTG(avg)_K}
   % \end{minipage} 
    }
\subfigure[]{
%\begin{minipage}{.32\textwidth}
    \centering
    \includegraphics[width=0.3\textwidth]{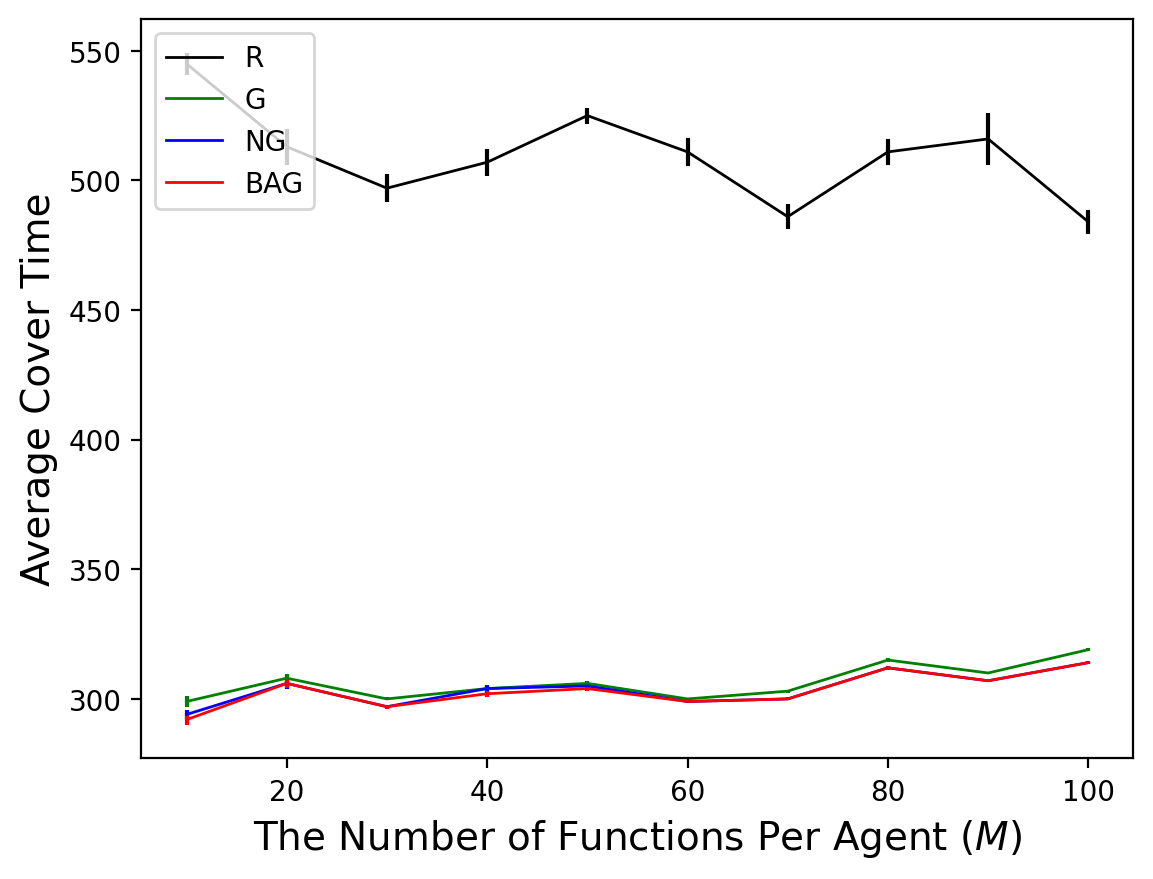}
  %  \caption{Runtime on Type Model instances with exponentially increasing noise}
    \label{fig:CTG(avg)_M}
    %\end{minipage}
    }
    %\caption{}
\subfigure[]{  
    %\begin{minipage}{.32\textwidth}
    \centering
    \includegraphics[width=0.3\textwidth]{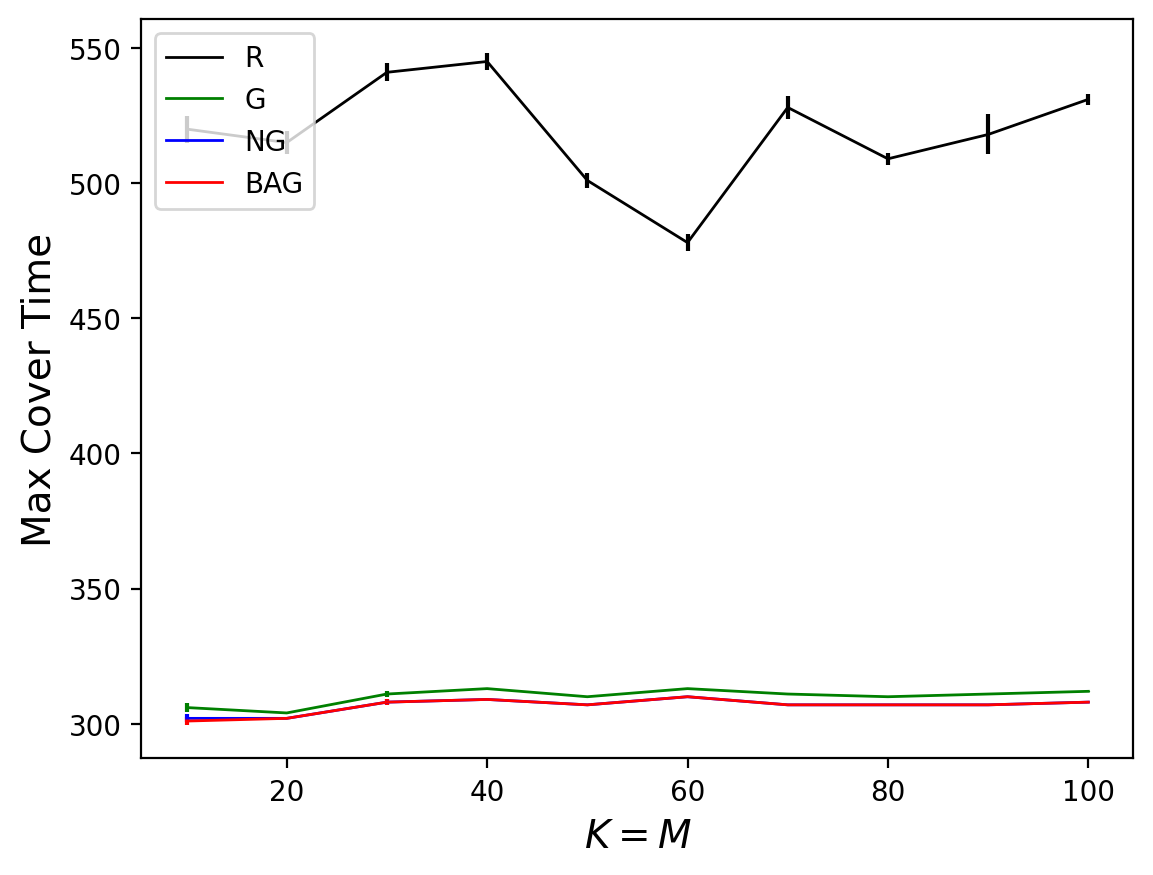}
 %   \caption{Runtime results for online setting on the type model}
    \label{fig:CTG(avg)_KM}
    %\end{minipage}
    }
    \caption{The algorithms' average cover times of agents on the CTG dataset when $K$ and $M$ vary. 
    }
    \label{fig:CTG(avg)}
\end{figure*}

In this section, we show the performance of algorithms on the PPPTS data set (\cref{fig:PPPTS}) and the CTG data set (\cref{fig:CTG}). The figures show the same trends. In addition, we also investigate the empirical results of these algorithms if the objective is the average cover time of agents, rather than the maximum cover time. We see that even in this case, our balanced adaptive greedy method still outperforms other algorithms on different data sets (\cref{fig:MFCC(avg)}, \cref{fig:PPPTS(avg)} and \cref{fig:CTG(avg)}).

\end{document}